\documentclass[pra,twocolumn,superscriptaddress]{revtex4}
\usepackage{graphicx}
\usepackage{amsmath}

\allowdisplaybreaks[1]

\begin{document}

\title{Towards loophole-free Bell inequality test with preselected \\
unsymmetrical singlet states of light}

\author{Magdalena Stobi\'nska}
\affiliation{Institute of Theoretical Physics and Astrophysics, University of Gda\'nsk, ul. Wita Stwosza 57, 80-952 Gda\'nsk, Poland}
\affiliation{Institute of Physics, Polish Academy of Sciences, Al.\ Lotnik\'ow 32/46, 02-668 Warsaw, Poland}
  
\author{Falk T\"oppel}  
\affiliation{Max Planck Institute for the Science of Light, Guenther-Scharowsky-Str. 1/Bldg. 24, 91058 Erlangen, Germany}
\affiliation{Institute for Optics, Information and Photonics, University of Erlangen-N\"urnberg, Staudtstr. 7/B2, 91058 Erlangen, Germany}

\author{Pavel Sekatski}
\affiliation{Group of Applied Physics, University of Geneva, Chemin de Pinchat 22, CH-1211 Geneva, Switzerland}

\author{Adam Buraczewski}
\affiliation{Faculty of Electronics and Information Technology, Warsaw University of Technology, ul. Nowowiejska 15/19, 00-665 Warsaw, Poland}

\begin{abstract}
Can a Bell test with no detection loophole be demonstrated for multi-photon entangled states of light within the current technology? We examine the possibility of a postselection-free CHSH-Bell inequality test with an unsymmetrical polarization singlet. To that end we employ a preselection procedure which is performed prior to the test. It allows using imperfect (coarse-grained) binary photodetection in the test. We show an example of preselection scheme which improves violation of the CHSH inequality with the micro-macro polarization singlet produced by the optimal quantum cloning. The preselection is realized by a quantum filter which is believed to be not useful for this purpose.
\end{abstract}

\maketitle

\section{Introduction}
 
Quantum mechanical laws apply to single particles, complex molecules involving tens of atoms as well as to living organisms~\cite{Arndt2009,Arndt1999,Brezger2002,Mohseni2008}. Recently, the first optical almost-loophole-free Bell tests for a two-photon singlet, eliminating the famous detection loophole, have been performed by the Zeilinger and Kwiat groups~\cite{Zeilinger,Kwiat}. Can this also be demonstrated for systems with higher mean photon numbers using the current technology? An indisputable Bell test~\cite{Bell,CHSH} ultimately rejects the local realistic description of the world in favor of quantum mechanics. It is also of practical importance allowing for implementation of quantum technology protocols such as device-independent quantum key distribution (QKD)~\cite{Acin2007}, randomness generation~\cite{Pironio2010} and reduction of communication complexity~\cite{Zukowski}.

The detection loophole arises from inefficient (lossy) photo-detection. The local realistic models do not necessarily satisfy the fair sampling assumption and they might exploit the postselection, i.e discarding some of the experimental data, to mimic the violation of a Bell inequality. Closing the detection loophole for a two-photon singlet was possible due to employment of the superconducting transition edge sensors~\cite{detectors1,detectors2}, the quantum detectors with a near-perfect efficiency. However, if we examined states of light involving large number of photons, elimination of this loophole would be more involved since the imperfect (coarse-grained) measurements come into play~\cite{Simon}.

Quantum phenomena on the macroscopic scale have been intriguing and puzzling to the physicists since the inception of the quantum theory. Recently, macroscopically populated entangled states of light became available experimentally: the micro-macro polarization singlet~\cite{DeMartini2008}, entangled bright squeezed vacuum~\cite{Macrobell,WitnesBSV}, and displaced single-photon path-entangled state~\cite{displacement,displacement-exp}. An important question of possibility of performing a loophole-free Bell test~\cite{A,B} for these state has been posed. The probability of no-detection event for these states is very low. This property gives hope to close the detection loophole. In Ref.~\cite{B} we showed that if the postselection is simply omitted, the micro-macro polarization state fails to pass the Bell test with efficient coarse-grained (binary) analog detection, although the loophole is closed. We also emphasized that preselection can solve this problem (it improves the visibility (distinguishability) of the multi-photon qubit in analog detection~\cite{Stobinska09}), but we did not provide any example to support our claim. However, the considerations in~\cite{Vitelli2010-2} contradict this statement: the authors showed that all so far tested preselections are not useful for increasing the distiguishability of the micro-macro polarization singlet in analog detection. Additionelly, the results in Refs.~\cite{Sekatski2010, Pomarico2011} emphasized the significance of the detection loophole for the test of macroscopic entanglement discussed  in Ref.~\cite{DeMartini2008}: it was demonstrated that in presence of this loophole separable states may falsely reveal entanglement.  Moreover, in Ref.~\cite{Simon} it was demonstrated that a single photon resolution is essential in observing the micro-macro entanglement with photon counting measurements.

Here, we examine a loophole(postselection)-free CHSH-Bell inequality test with preselected unsymmetrical polarization singlet states of light of a general form and imperfect binary analog detection. We explicitly show an example of preselection scheme which improves violation of the CHSH inequality with the micro-macro polarization singlet produced by the optimal quantum cloning.

In the unsymmetrical singlets under consideration, one of the modes is occupied by a single photon (the micro-qubit), whereas the second one contains a pair of mutually orthogonal multi-photon states (the multi-photon qubit). We assume that the average photon number in the multi-photon qubit can be controlled by some external parameter in an experimental setup, and it may vary from a single photon to the macroscopic quantity of thousands of photons. Furthermore, we consider a Bell test based only on linear optical elements. The unsymmetrical singlet is prepared before the test by a special filtering procedure applied to the mode containing the multi-photon state. The filter is described by a POVM (positive operator valued measure). For example, it may be realized by the modulus of intensity difference filter~\cite{Stobinska2011,NJP2014}. Filtering belongs to the conditional state preparation, not to the test. Only if the state is successfully preselected, the Bell test is performed where every measurement outcome is conclusive and is taken into account. This eliminates the necessity of data postselection and closes the detection loophole.

This paper is organized as follows. In Section~\ref{sec2} we discuss a general scenario of the CHSH-Bell inequality test with preselection strategy for an unsymmetrical singlet. Section~\ref{sec3} is devoted to a short summary of the experimentally available unsymmetrical polarization singlets of light. We further discuss the CHSH inequality violation for these states preselected by the special case of the modulus of intensity difference filter, namely the corner filter, in Section~\ref{sec4}. Finally, we discuss the possible future steps towards genuine loophole-free Bell test for states of light with large photon population.

\section{CHSH-Bell test with preselection} \label{sec2}

\begin{figure}
  \begin{center}
    \includegraphics[height=3cm]{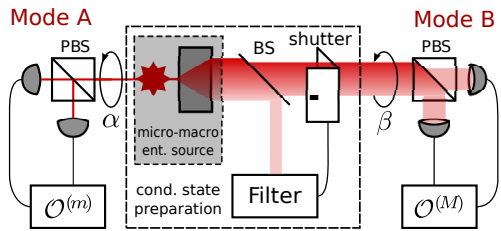}\\
   \end{center}
  \caption{Bell inequality test involving a preselection strategy and un unsymmetrical singlet state of light.}           
 \label{schema}
\end{figure} 

The Bell inequality test with an unsymmetrical singlet state and imperfect intensity measurements has to employ a preselection strategy~\cite{Stobinska09,B}. The role of preselection is to prevent from deterioration of the ability to witness quantum correlations in the singlet state, resulting from coarse-graining measurements. We call the singlet unsymmetrical if the dimensions of the Hilbert spaces corresponding to its modes are unequal. 

Let us start our discussion with a two-mode unsymmetrical polarization singlet state of a general form
\begin{equation}
|\Psi^-\rangle=1/\sqrt{2}(|1_{\varphi}\rangle_A |\bar{\Phi}\rangle_B - |1_{\varphi_{\perp}}\rangle_A |\Phi\rangle_B)
\label{micro-macro-singlet1}
\end{equation}
and an arbitrary preselection strategy executed by a POVM $\mathcal{P}$. The states $|1_{\varphi}\rangle$ and $|1_{\varphi_{\perp}}\rangle$ denote a micro-qubit, e.g. a single photon in polarization $\varphi$ and $\varphi_{\perp}$ respectively, whereas $|\Phi\rangle$ and $|\bar{\Phi}\rangle$ are multi-photon states which constitute a multi-photon qubit. Of course, the two states of the qubits are pairwise orthogonal $\langle 1_{\varphi}|1_{\varphi_{\perp}}\rangle =0$, $\langle\Phi|\bar{\Phi}\rangle=0$.

A setup for the Bell test is depicted in Fig.~\ref{schema}. The multi-photon part of the state (mode B) impinges on a beam splitter (BS) with a low reflectivity $r$, e.g. $10\%$, which taps only a small fraction of the state leaving it almost unaffected. Next, the preselection strategy is implemented by the analysis of the reflected part: it is examined by a filter described by a POVM $\mathcal{P}$ and the result is feed forwarded to the transmitted beam. This procedure belongs to conditional state preparation before the test. After the successful preselection, the Bell test consists of polarization rotations, by the angles $\alpha$, $\alpha'$ on mode $A$ and $\beta$, $\beta'$ on mode $B$, and intensity measurements of polarization components of both modes.

In general, the operator $\mathcal{P}$ may suffer from lack of the rotational invariance being an important property of the original singlet and thus, the form of the preselected state may be basis dependent. For example, for the modulus of intensity difference filter it is known that it improves the visibility of a multi-photon state for a measurement in one polarization basis but deteriorates for measurements in the other polarization bases. Thus, such filtering strategies are believed to be useless for preselection~\cite{Vitelli2010-2}. This problem arises if the usual settings for the CHSH-Bell inequality are considered: $\alpha = 0$, $\alpha' = \pi/2$, $\beta = -\pi/4$, $\beta' = \pi/4$.  However, it does not need to be the case if nonorthogonal polarization directions in measurements on the multi-photon mode are chosen, at the expense of obtaining a non-maximal Bell inequality violation. Moreover, the visibility is not the only factor contributing to the CHSH-Bell parameter computed for an unsymmetrical singlet. The other parameter is a quantity which we call antivisibility and we explain its physical meaning below. In order to maximize the value of the Bell violation, the rotation angles on the multi-photon mode should optimize the two parameters simultaneously.

We consider the CHSH inequality with Bell parameter
\begin{equation}
\label{Bell}
B = E(\alpha, \beta) + E(\alpha, \beta') + E(\alpha', \beta) - E(\alpha', \beta')
\end{equation}
and $E(\alpha, \beta) = \langle \mathcal{O}^{(m)}(\alpha)\otimes\mathcal{O}^{(M)}(\beta) \rangle$ is the correlation function where one observer, Alice, measures the microscopic part (mode A) and the other, Bob, measures the multi-photon part of the singlet (mode B).  We assume the ideal measurement operator $\mathcal{O}^{(m)}(\alpha)= \lvert 1_{\alpha}\rangle\langle 1_{\alpha}\rvert- \lvert 1_{\alpha_{\perp}}\rangle\langle 1_{\alpha_{\perp}}\rvert$ for the microscopic qubit.  Rotating the polarization of the microscopic part by an angle $\alpha$ yields
\begin{align*}
\lvert 1_{\alpha}\rangle ={}& \cos(\alpha/2)\lvert 1\rangle + \sin(\alpha/2)\lvert 1_{\perp}\rangle,
\\
\lvert 1_{\alpha_{\perp}}\rangle ={}&-\sin(\alpha/2)\lvert 1\rangle + \cos(\alpha/2)\lvert 1_{\perp}\rangle,
\end{align*}
which allows expressing the micro observable in terms of the projectors in the reference basis $\varphi=0$
\begin{align}
\mathcal{O}^{(m)}(\alpha) ={}& \cos \alpha \left(|1\rangle \langle 1| - |1_{\perp}\rangle \langle 1_{\perp}| \right)
\\
{}+{}& \sin \alpha \left(|1\rangle \langle 1_{\perp}| + |1_{\perp}\rangle \langle 1| \right). 
\nonumber
\end{align}
For the multi-photon mode we take the binary threshold detection operator $\mathcal{O}^{(M)}(\beta)$ adapted to the preselection strategy $\mathcal{P}$. The value $+1$ ($-1$) is assigned to this observable when the state $|\Phi\rangle$ ($|\bar{\Phi}\rangle$) is identified. We assume it belongs to a class of diagonal observables such that $\mathrm{Tr}\{\mathcal{O}^{(M)}(\beta)\}=0$.  The general form of such observable reads
\begin{equation}
  \label{eq:general_OM}
  \mathcal{O}^{(M)}(\beta) =
   \sum_{\substack{k,l=0\\C(k,l)}}^{\infty}
   |k_{\beta}, l_{\beta_{\perp}}\rangle \langle k_{\beta}, l_{\beta_{\perp}}|,
\end{equation}
where the condition $C(k,l)$ is such that it ensures the observable to be traceless.

After a short algebra, we obtain the correlation function for the state in Eq.~(\ref{micro-macro-singlet1})
\begin{equation}
E(\alpha, \beta) = -\cos \alpha\, V^{\theta} (\beta) - \sin \alpha \, A^{\theta} (\beta),
\label{correlation}
\end{equation}
where $V^{\theta} (\beta) = \langle \Phi^{\theta}|\mathcal{O}^{(M)}(\beta)|\Phi^{\theta}\rangle$ is the visibility and  $A^{\theta} (\beta) = \langle \bar{\Phi}^{\theta}|\mathcal{O}^{(M)}(\beta)|\Phi^{\theta}\rangle$ is called the antivisibility of the state $\Phi$ preselected in the polarization basis $\theta$ and observed in the polarization basis $\beta$. The antivisibility quantifies the ability of the observable $\mathcal{O}^{(M)}(\beta)$ to erase the information on which state of $\lvert \bar \Phi_\theta\rangle$ or $\lvert \Phi_\theta\rangle$ entered the detector. In this derivation, due to the condition $\mathrm{Tr}\{\mathcal{O}^{(M)}(\beta)\}=0$, we noticed that $V^{\theta}_{\perp} (\beta) = \langle \bar{\Phi}^{\theta}|\mathcal{O}^{(M)}(\beta)|\bar{\Phi}^{\theta}\rangle = -V^{\theta}(\beta)$.  Without the loss of generality, we also took that $A^{\theta} (\beta)$ is real-valued. After inserting Eq.~(\ref{correlation}) into Eq.~(\ref{Bell}) we obtain
\begin{align}
\label{eq:Bell_any_angles}
B ={}& -\cos \alpha\, (V^{\theta} (\beta) + V^{\theta} (\beta') ) - \sin \alpha \, (A^{\theta} (\beta) + A^{\theta} (\beta') )\nonumber\\
&{}-\cos \alpha'\, (V^{\theta} (\beta) - V^{\theta} (\beta') ) - \sin \alpha' \, (A^{\theta} (\beta) - A^{\theta} (\beta') ).\nonumber\\
\end{align}

We will first consider the following rotation angles $\theta=0$ for preselection and $\alpha=0$, $\alpha'=\pi/2$, $\beta'=-\beta$ for the Bell test. This choice is quite natural because in the limit of a small photon population in mode $B$, i.e. for a two-photon singlet, these are the optimal angles maximizing the value of the CHSH-Bell parameter with $\beta= -\pi/4$.  Later, we will show that for the specific examples we examined it is also the optimal set of angles even for amplified mode $B$ however, the optimal value of $\beta$ ($\beta_{\text{opt}}$) changes with the mean number of photons in multi-photon qubit in presence of preselection. In this case, the Bell parameter reads
\begin{equation}
B_{\text{opt}}= -\left(V (\beta_{\text{opt}})+V (-\beta_{\text{opt}}) + A(\beta_{\text{opt}}) - A(-\beta_{\text{opt}})\right).
\nonumber
\end{equation}
Due to the diagonal form of the multi-photon observable, the visibility and antivisibility can be expressed as a convex sum of contributions resulting from various photon-number sectors
\begin{align}
V(\beta)=&\sum_{k=0}^{\infty}\beta_k^2\,V_k(\beta),&
A(\beta)=&\sum_{k=0}^{\infty}\beta_k^2\,A_k(\beta),
\label{eq:decomposition}
\end{align}
where $V_k(\beta)$ and $A_k(\beta)$ are computed for the $k$th photon-number sector of a multi-photon qubit (we can always decompose multi-photon states in the Fock basis as follows $|\Phi\rangle =\sum_{k=0}^{\infty}\beta_k\,\lvert\Phi_k\rangle$ with $\lvert\Phi_k\rangle = \sum_{j=0}^k \xi_{k,j} |k-j, j_{\perp}\rangle$ where $\beta_k$ and $\xi_{k,j}$ are certain probability amplitudes). Thus, similar decomposition holds true for the Bell parameter 
\begin{align}
\label{eq:Bell_decomposition}
B=&\sum_{k=0}^{\infty}\beta_k^2\,B_k, \\
B_k=&-\left(V_k(\beta_{\text{opt}})+V_k (-\beta_{\text{opt}}) + A_k(\beta_{\text{opt}}) - A_k(-\beta_{\text{opt}})\right).\nonumber
\end{align}
Decompositions in Eqs.~(\ref{eq:decomposition}) and (\ref{eq:Bell_decomposition}) give insight into the contribution of each sector separately by taking into account the structure of the multi-photon qubit. Due to that we know which photon numbers lead to the Bell violation most and which deteriorate it.

The above formula may be further simplified by noticing that when $\xi_{k,j}$ and $\bar{\xi}_{k,j}$ fulfill additional conditions, e.g.\ $\xi_{k,j}=0$ and $\bar{\xi}_{k,j}\not=0$ for odd $j$ but $\xi_{k,j}\not=0$ and $\bar{\xi}_{k,j}=0$ for even $j$, then $V_k (\beta)=V_k (-\beta)$ and $A_k(\beta) = -A_k(-\beta)$ (see Appendix~A). In this case it is possible to write Eq.~(\ref{eq:Bell_decomposition}) as $B_k=-2\left(V_k(\beta_{\text{opt}})+ A_k(\beta_{\text{opt}})\right)$.

\section{Example: Unsymmetrical polarization singlet states of light emerging from phase-covariant quantum cloner}\label{sec3}

In this section we will discuss a specific example of the experimentally available unsymmetrical polarization singlet states of light. They are produced in the process of the phase-covariant optimal quantum cloning. It is based on phase sensitive parametric amplification~\cite{Sekatski2010,DeMartini-PRL,Masha} and requires a pair of linearly polarized photons in a standard singlet state, obtained through parametric down conversion, as an input. The single photon seeding is coherently amplified to produce a multi-photon state, by an intensely pumped high gain g nonlinear medium (the cloner). The equatorial states of the Poincar\'e sphere of all polarization states, parametrized by the polar angle $\varphi \in \langle 0,2\pi)$, are privileged for the phase-covariant cloners. Only for this subspace their Hamiltonian $H = \frac{i\chi}{2}\left({a_{\varphi}^{\dagger}}^2 + {a_{\varphi_{\perp}}^{\dagger}}^2 \right) + \mathrm{h.c.}$ is rotationally invariant and they work equally well for all the equatorial states. The operators $a_{\varphi}^{\dagger}$ and ${a_{\varphi_{\perp}}^{\dagger}}$ denote the creation operators for the equatorial polarization modes $\varphi$ and $\varphi_{\perp}$, respectively and $\chi$ is the coupling strength, proportional to the pumping power. We restrict ourselves to the equatorial polarization state subspace for the seeding photon. The subspace basis is set by two states, $|1_{\varphi}\rangle = 1/\sqrt{2}(|1_H\rangle + e^{i\varphi}|1_V\rangle)$ and its orthogonal counterpart $|1_{\varphi_{\perp}}\rangle $, where $|k_H\rangle$ ($|l_V\rangle$) denote $k$ ($l$) photons polarized horizontally (vertically) and $\varphi_{\perp} = \varphi + \pi$. Due to its rotational invariance, we express the initial singlet in this basis $|\psi^-\rangle=1/\sqrt{2}(|1_{\varphi}\rangle_A |1_{\varphi_{\perp}}\rangle_B - |1_{\varphi_{\perp}}\rangle_A |1_{\varphi}\rangle_B)$. Cloning is a unitary process and the original two-photon entanglement is transferred to the unsymmetrical singlet with  
\begin{align}
|\Phi\rangle ={}& \sum_{i,j=0}^{\infty} \!\gamma_{ij} \big|(2i+1)_{\varphi},(2j)_{\varphi_{\perp}}\rangle,
\label{DeMartini}\\
|\bar{\Phi}\rangle ={}& \sum_{i,j=0}^{\infty} \!\gamma_{ij} \big|(2j)_{\varphi},(2i+1)_{\varphi_{\perp}}\rangle\nonumber.
\end{align}
$|\Phi\rangle$ and $|\bar{\Phi}\rangle$ are the amplified single photons, with the real-valued probability amplitude $\gamma_{ij}=\cosh^{-2}g \left(\tanh g /2\right)^{i+j} \sqrt{(1+2i)!(2j)!}/i!/j!$ where $g$ is the parametric gain. In the experiment, their average population equals $4\sinh^2g+1$, varied from less than one up to $10^{4}$ of photons. Due to different parity of occupation number in the Fock state basis, $|\Phi\rangle$ and $|\bar{\Phi}\rangle$ are orthogonal. However, in high photon number regime photon detectors are not single photon resolving~\cite{Masha} and visibility of the multi-photon states is quite low~\cite{multiporty,CPC}, making them inapplicable for quantum protocols and Bell inequality test.

\subsection{Quantum filtering}

Visibility of the multi-photon qubit can be improved by quantum state filtering performed by certain POVM filters. They modify the state but preserve quantum superpositions. Recently, such a filter has been proposed~\cite{Stobinska2011}: the modulus of intensity difference filter selects two-mode states of light whose mode populations differ by more than a certain threshold $\delta_{\text{th}}$.  It estimates the absolute value of difference instead of the difference. It does not provide any information on which polarization mode was more populated and thus, it is not able to distinguish the multi-photon states and preserves superpositions. It performs almost a nondestructive measurement implemented by the tapping and feed forward loop, i.e. the filtered output state is almost pure. Qualitatively, it approximates the following operation
\begin{equation}
  \mathcal{P}_{\text{MDF}} = \kern-1em\sum_{\substack{k,l=0\\\lvert k-l\rvert\ge {\delta}_{\text{th}}}}^{\infty}\kern-1em |k,l\rangle \langle k,l|,
  \label{theoretical-preselection}
\end{equation} 
where $|k,l\rangle$ is a polarization two mode Fock state. Below, we will briefly discuss its physical implementation and the principle of operation. More detailed discussion, including action of the filter on a multi-photon states from Eq.~(\ref{DeMartini}), is given in~\cite{Stobinska2011} and description of the first attempt of its experimental realization can be found in~\cite{NJP2014}.

\begin{figure}
  \begin{center}
    \includegraphics[height=4cm]{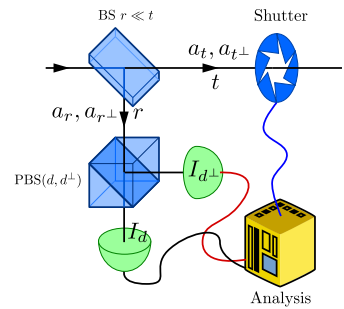}\\
   \end{center}
  \caption{Realization and application of the modulus of intensity difference and other quantum filters.}           
 \label{filter}
\end{figure}
The approximate realization and application of the quantum filter is presented in Fig.~\ref{filter}. The state to be filtered enters a feed forward loop. It impinges on a highly unbalanced beam splitter with a very small reflectivity $r$ which splits the state into the reflected $a_r$ ($a_{r\perp}$) and transmitted modes $a_t$ ($a_{t\perp}$). The reflected mode is examined by the quantum filter. Since the reflected and transmitted beams are correlated, estimating the modulus of the population difference for the former gives an estimate for the latter. Depending on the result of this analysis, the transmitted mode is either passed or blocked by the shutter. In this way, it is possible to block light of unwanted properties. The quantum filter consists of a polarization beam splitter (PBS), which works in a basis unbiased with respect to the polarization basis of the incoming field ($a_{d(\perp)} = 1/\sqrt{2}(a_r\pm a_{r\perp})$), and the photon counting detectors. These may be the superconducting transition-edge sensors (TESs) with a very high quantum efficiency of ca. $95\%$~\cite{detectors1,detectors2}. In the up-to-date experiments they achieved the overall efficiency of $75\%$~\cite{Zeilinger,Kwiat}. This result can be improved by using the integrated optics setups. These detectors posses a single photon resolution in the range of ca. $0-23$ of photons. This makes quantum filtering of the macroscopically populated superpositions of light experimentally feasible, taking into account that the population of the reflected mode may constitute, say, $1-10\%$ of the total population of the incoming field. In the higher photon number range the detectors may also work quite well for the filtering purposes, since the relative error of the measurement (the uncertainty in photon counting compared to the incoming population) is pretty small: for $1000$ incoming photons it is ca. $30$ photons.

The key property of the filter is that the more unequally populated a two-mode Fock state, entering the PBS, is, the more equally populated are the output modes the more the measured output modes are roughly equally populated (and vice versa). This effect is especially pronounced for highly populated states. It allows with high probability to estimate the population difference in the reflected mode, and in consequence in the transmitted mode as well.

\section{CHSH-Bell test for the unsymmetrical polarization singlet emerging from phase-covariant quantum cloner}\label{sec4}

In order to investigate the Bell inequality violation for the unsymmetrical singlet described in Section III, we re-write the multi-photon states given in Eq.~(\ref{DeMartini}) as superpositions of states of a fixed photon number $2k+1$ distributed over the two polarization modes
\begin{align}
\label{sectors}
\lvert\Phi\rangle ={}&\sum_{k=0}^{\infty}\beta_k\,\lvert\Phi_k\rangle, \lvert\Phi_k\rangle = \tfrac{1}{\sqrt{\mathcal{N}_k}}
  \left({a^{\dagger}}^2+{\left.a_{\perp}^{\dagger}\right.}^2\right)^k a^{\dagger}\,\lvert 0\rangle, 
\\
\lvert\bar{\Phi}\rangle ={}&\sum_{k=0}^{\infty}\beta_k\lvert{\bar{\Phi}_{k}}\rangle, \lvert\bar{\Phi}_{k}\rangle = \tfrac{1}{\sqrt{\mathcal{N}_k}}\,
  \left({a^{\dagger}}^2+{\left.{a}_{\perp}^{\dagger}\right.}^2\right)^k\, {a}_{\perp}^{\dagger}\,\lvert 0\rangle, \nonumber
\end{align}
where we took $\varphi=0$, $\mathcal{N}_k=4^k\,{k!}^2\,(1+k)$, 
\begin{equation}
\beta_k=\cosh^{-2}g\,(\tanh g)^k\,\sqrt{1+k}, \quad \sum_{k=0}^{\infty}\beta_k^2=1.
\label{betanop}
\end{equation} 

We note that the visibility and antivisibility have the following symmetry properties for states from Eq.~(\ref{sectors}): $V(\beta)=V(-\beta)$, $A(\beta) = - A(-\beta)$, see Eq.~(\ref{Vk})-(\ref{Ak}) and Appendix~A.  They allow to simplify the Bell parameter to the following form
\begin{equation}
\label{eq:Bell_sum}
B=-2(V(\beta_{\text{opt}})+A(\beta_{\text{opt}})).
\end{equation}
It depends on two factors: the visibility $V(\beta_{\text{opt}})$ and the antivisibility $A(\beta_{\text{opt}})$ measured in the same basis, rotated by $\beta_{\text{opt}}$ with respect to the reference one. A violation of Bell's inequality is obtained if $|V(\beta_{\text{opt}}) + A(\beta_{\text{opt}})| >1$. We emphasize that this result holds true for any preselection strategy applied to states in Eq.~(\ref{sectors})
\begin{equation}
\label{eq:Bell_sum_P}
B^P=-2(V^P(\beta_{\text{opt}})+A^P(\beta_{\text{opt}})),
\end{equation}
where the index $P$ denotes the quantities evaluated for the preselected multi-photon states.

\subsection{Bell test without preselection}

In order to analyze the difficulties with Bell inequality violation we adapt the following observable
\begin{equation}
  \mathcal{O}^{(M)}(\beta) =
    \left(
   \sum_{\substack{k,l=0\\k-l\ge0}}^{\infty}
   -
   \sum_{\substack{k,l=0\\k-l<0}}^{\infty}
   \right)
   |k_{\beta}, l_{\beta_{\perp}}\rangle \langle k_{\beta}, l_{\beta_{\perp}}|.
\label{eq:macro_observable}
\end{equation}
This is a diagonal, traceless operator. It is well-suited to the photon number distribution of the multi-photon qubit in Eq.~(\ref{sectors}): $\lvert\Phi\rangle$ and $\lvert\bar{\Phi}\rangle$ have unequal average population in the two polarization modes. The mean photon number in polarization $\varphi$ is three times larger than the mean number of photons in polarization $\varphi_{\perp}$ in $\lvert\Phi\rangle$ (the opposite relation holds true for $\lvert\bar{\Phi}\rangle$)~\cite{SciarrinoWigner}. This property allows to distinguish these states in analog detection. Possible implementation of the measurement of the observable $\mathcal{O}^{(M)}(\beta)$ would require splitting the two-mode polarization multi-photon beam by a polarization beam splitter followed by TESs detection. We would assign $+1$ to the measurement outcome if the signal difference between polarizations $\beta$ and $\beta_{\perp}$ was positive and $-1$ if it was negative. The detectors composed of PIN diodes followed by charge-sensitive amplifiers, such as those used in~\cite{Stokes} for measuring uncertainties of the Stokes variables for macroscopically populated squeezed vacuum, should also be sufficient for implementation of this measurement adequately.

The visibility and antivisibility evaluated for this observable equal
\begin{align}
\label{Vk}
V_k(\beta)=& \dfrac{1}{\mathcal{N}_k}
\begin{aligned}[t]
&\bigg[\sum_{u-w\ge 0} - \sum_{u-w<0}\bigg]\dfrac{\delta_{u+w,2k+1}}{u!\,w!}\\
&\Bigg\{\sum_{j=0}^k\binom{k}{j} (2j+1)!\,(2k-2j)!\\
&\quad\sum_{m=0}^u \sum_{n=0}^w \binom{u}{m}\binom{w}{n}(-1)^{n}\\
&\qquad\cos^{u-m}\big(\tfrac{\beta}{2}\big) \sin^{m}\big(\tfrac{\beta}{2}\big)\\
&\qquad\sin^{w-n}\big(\tfrac{\beta}{2}\big) \cos^{n}\big(\tfrac{\beta}{2}\big)\delta_{2k-2j,m+n}\Bigg\}^2,
\end{aligned}\kern-3em
\end{align}

\begin{align}
\label{Ak}
A_k(\beta)
={}& 
\begin{aligned}[t]
&\dfrac{1}{\mathcal{N}_k}\bigg[\sum_{u-w>0} - \sum_{u-w<0}\bigg]\dfrac{\delta_{u+w,2k+1}}{u!\,w!}\\
&\Bigg\{\sum_{j=0}^k \binom{k}{j} (2j+1)!\,(2k-2j)!\\
&\quad \sum_{m=0}^u \sum_{n=0}^w \binom{u}{m}\binom{w}{n}(-1)^{n}\\
&\qquad\cos^{u-m}\big(\tfrac{\beta}{2}\big) \sin^{m}\big(\tfrac{\beta}{2}\big)\\
&\qquad\sin^{w-n}\big(\tfrac{\beta}{2}\big) \cos^{n}\big(\tfrac{\beta}{2}\big)\delta_{2k-2j,m+n}\Bigg\}\\
\cdot&\Bigg\{\sum_{j=0}^k \binom{k}{j} (2j)!\,(2k+1-2j)!\\
&\quad \sum_{m=0}^u \sum_{n=0}^w \binom{u}{m}\binom{w}{n}(-1)^{n}\\
&\qquad \cos^{u-m}\big(\tfrac{\beta}{2}\big) \sin^{m}\big(\tfrac{\beta}{2}\big)\\
&\qquad \sin^{w-n}\big(\tfrac{\beta}{2}\big) \cos^{n}\big(\tfrac{\beta}{2}\big)\delta_{2k+1-2j,m+n}\Bigg\},
\end{aligned}\kern-3em
\end{align}
where $\delta_{i,j}$ denotes the Kronecker's delta, equal 1 when $i=j$ and 0 otherwise.

We now look for $\beta_{\text{opt}}$ which maximizes the value of the CHSH-Bell parameter in Eq.~(\ref{eq:Bell_sum}). We numerically checked that regardless the mean number of photons in the multi-photon state, $\beta_{\text{opt}}=-\pi/4$. 

\begin{figure}
\begin{flushright}
\includegraphics[height=4cm]{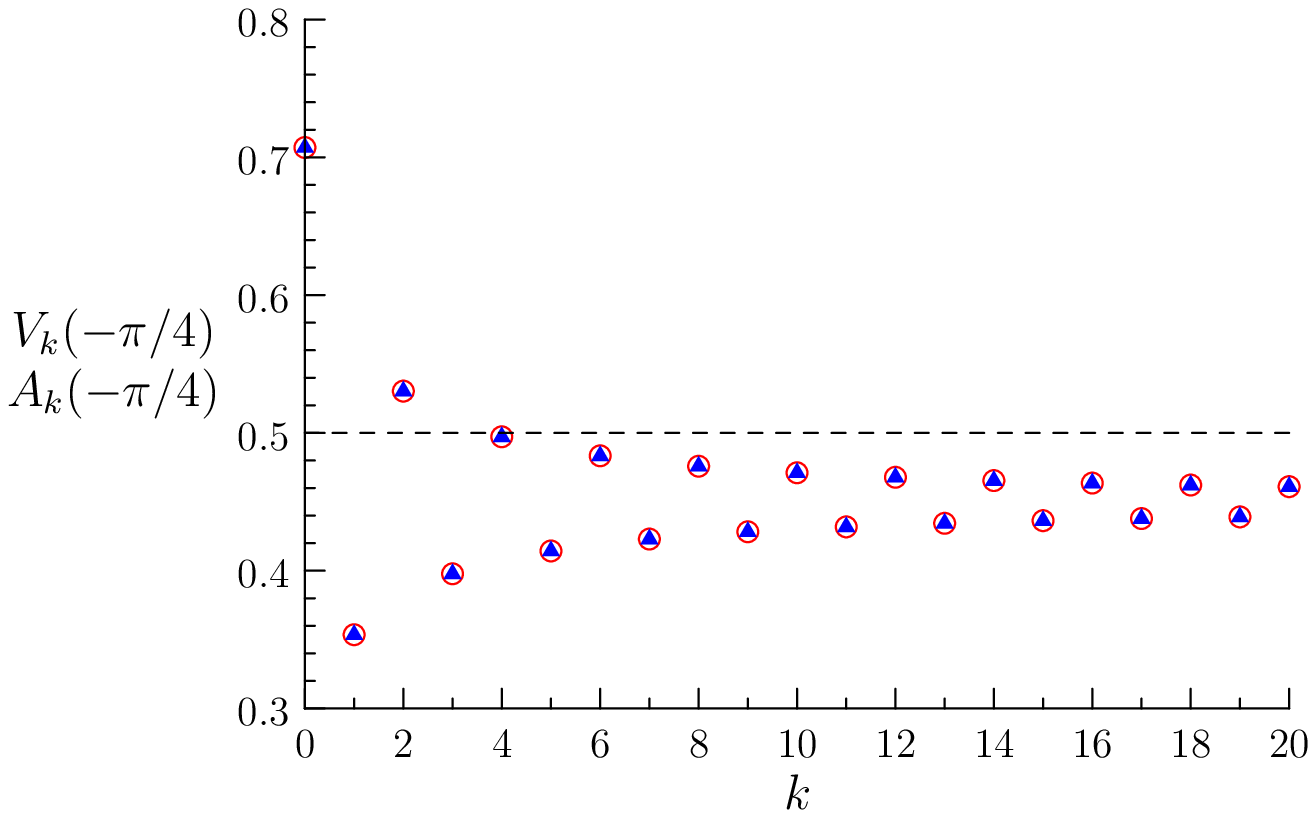}\kern5em\ \\
\includegraphics[height=4cm]{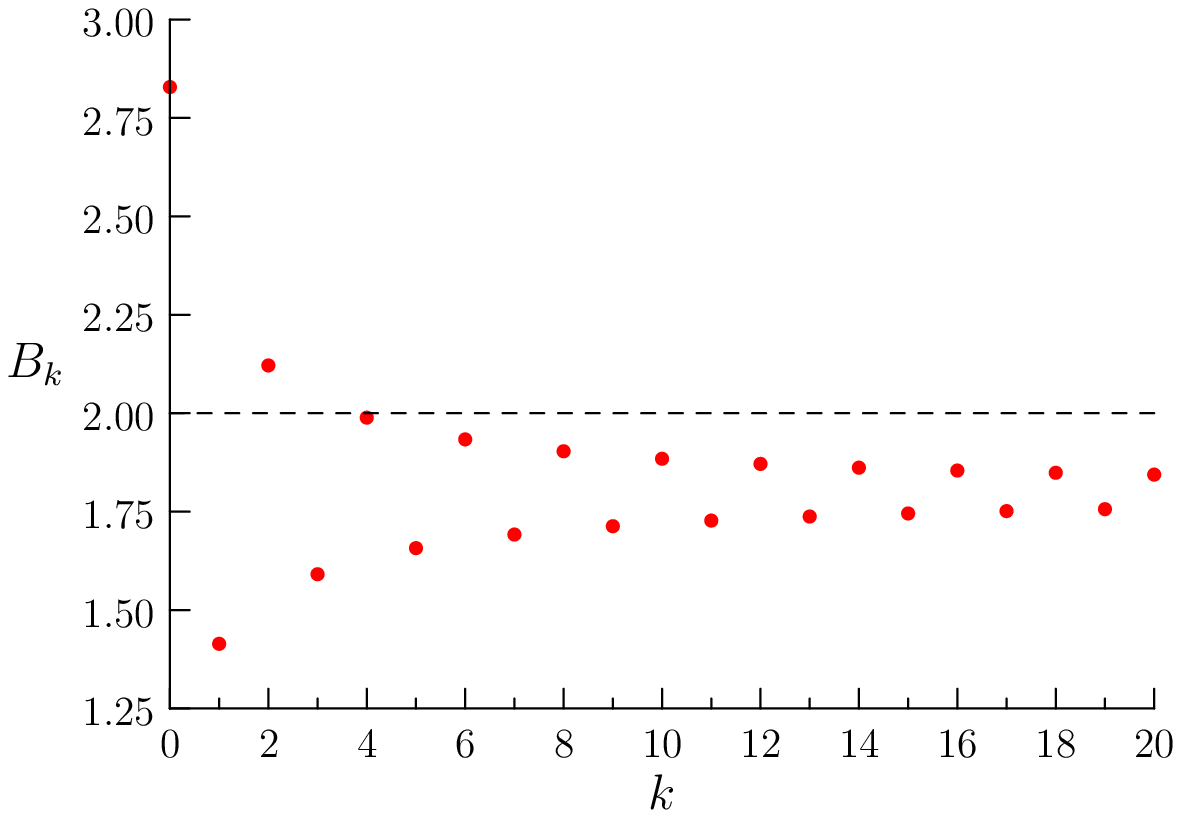}\kern5em\ \\
\end{flushright}
\caption{Plot of the visibility $V_k$ from Eq.~(\ref{Vk}) (top figure, red circles), antivisibility $A_k$ from Eq.~(\ref{Ak}) (top figure, blue triangles) and the Bell parameter $B_k$ from Eq.~(\ref{eq:Bell_decomposition}) (bottom figure) computed for $\beta_{\text{opt}}=-\pi/4$ as a function of $k$ corresponding to $(2k+1)$-photon-number sector of the multi-photon states $\lvert\Phi_k\rangle$ and $\lvert\bar{\Phi}_{k}\rangle$ given in Eq.~(\ref{sectors}). (Color online)}
\label{fig:nopreselection}
\end{figure}

\begin{figure}
\begin{flushright}
\includegraphics[height=4cm]{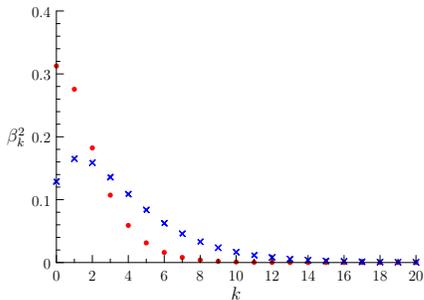}\kern5em\ \\
\end{flushright}
\caption{Plot of $\beta_k^2$ defined in Eq.~(\ref{betanop}) for $g=0.8$ (red dots) and $g=1.1$ (blue crosses). (Color online)}
\label{fig:nopreselection_beta}
\end{figure}

In Fig.~\ref{fig:nopreselection} we depicted the visibility, the antivisibility and the Bell parameter for each $2k+1$-photon-number sector separately, computed for the optimal set of angles.  We note that $V_k$ and $A_k$ both quickly tend to the value less than $1/2$ for increasing $k$ and the Bell parameter $B_k$ drops below 2 for $k\ge 3$. The probability distribution $\beta_k^2$ depends on the amplification gain $g$ and gives a weight to each $B_k$ contributing to the Bell parameter $B$. Since the greatest weight is given to the small values of $k$, see Fig.~\ref{fig:nopreselection_beta}, it is possible to weakly violate the CHSH-Bell inequality for a very small average photon population, so for very small $g$. Note that the considered state approaches the Bell singlet state in the limit $g \to 0$. For $g=0.8$ (total mean number of photons equals $4.15$) $B=2.06$, but for $g=1.1$ (total mean number of photons equals $8.13$) we found $B=2.01$. Thus, Bell inequality violation for the multi-photon unsymmetrical singlet with the multi-photon qubit given in Eq.~(\ref{sectors}) and the population of already few photons on average, is difficult to detect.

\subsection{Bell test with a preselection strategy followed by imperfect binary detection}

A preselection strategy described by a POVM of the form
\begin{equation}
  \label{pre}
  \mathcal{P}_{\text{C}}=\kern-2em\sum_{\substack{k,l=0\\\text{C}(\sigma=k+l,\Delta=l-k)}}^{\infty}\kern-2em
  \lvert k,l\rangle\langle k,l\rvert,
\end{equation}
with a general preselection condition $\text{C}$ on $\sigma$ and $\Delta$, insignificantly influences the form of the convex sum in which the Bell parameter is expressed in Eq~(\ref{eq:decomposition}). The detailed calculations are presented in Appendix~B. Assuming that the condition $\text{C}(\sigma,\Delta)$ is symmetric, i.e. $\text{C}(\sigma,\Delta)=\text{C}(\sigma,-\Delta)$, we show that the preselection modifies both multi-photon states from Eq.~(\ref{sectors}) in the same way
\begin{align}
\label{preselected_multiqubit}
  \lvert\Phi_k^P\rangle=&
  \tfrac{1}{\sqrt{\mathcal{N}_k^P}}\,\mathcal{P}_{\text{C}}\,\lvert\Phi_k\rangle, \, \lvert\bar{\Phi}_k^P\rangle= \tfrac{1}{\sqrt{\mathcal{N}_k^P}}\,
  \mathcal{P}_{\text{C}}\,\lvert\bar{\Phi}_k\rangle, \\
 \lvert\Phi^P\rangle=&\sum_{k=0}^{\infty}\beta_k^P\,\lvert\Phi_k^P\rangle, \, \lvert\bar{\Phi}^P\rangle=\sum_{k=0}^{\infty}\beta_k^P\lvert\bar{\Phi}_k^P\rangle,
 \nonumber
\end{align}
where $\mathcal{N}_k^P$ (Eq.~\ref{norm1}) is the new normalization constant and $\beta_k^P$ (Eq.~(\ref{beta})) is the new probability amplitude in the decomposition of the states into the photon-number sectors. 

We would like to mention that although preselection may enable application of coarse-graining measurements in observing quantum effects in the multi-photon superpositions by increasing their visibility, e.g. in Bell test, it will not be a direct remedy to the deteriorating effect of losses in an experimental setup. Preselection helps in conditional state generation. Thus, the robustness of the new (preselected) state will determine the robustness of the whole Bell test against losses.

\subsection{Example: the corner filter}

\begin{figure}
  \begin{tabular}{cc}
    \raisebox{2cm}{(a)}
    \includegraphics[height=3cm]{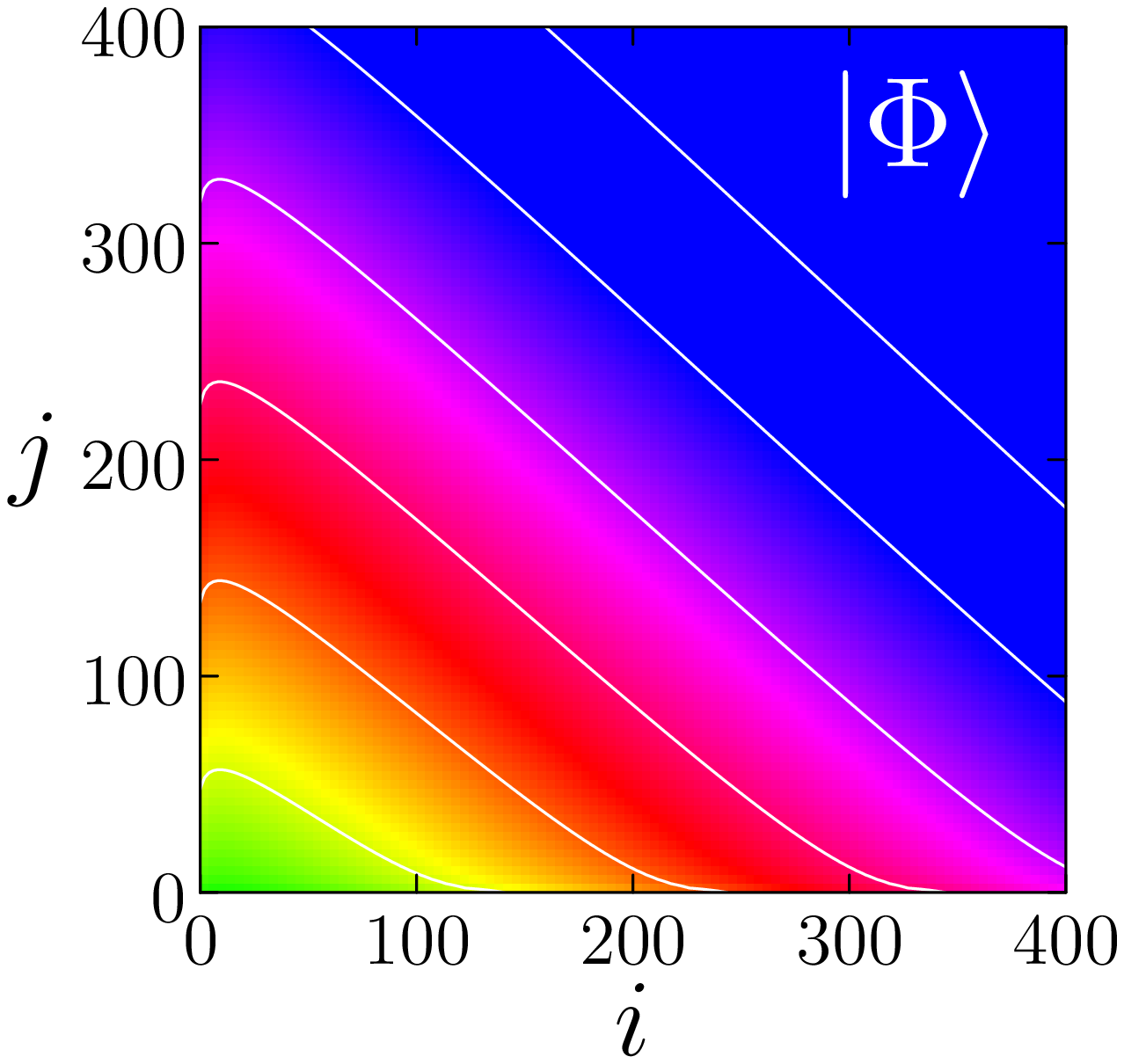}&
    \includegraphics[height=3cm]{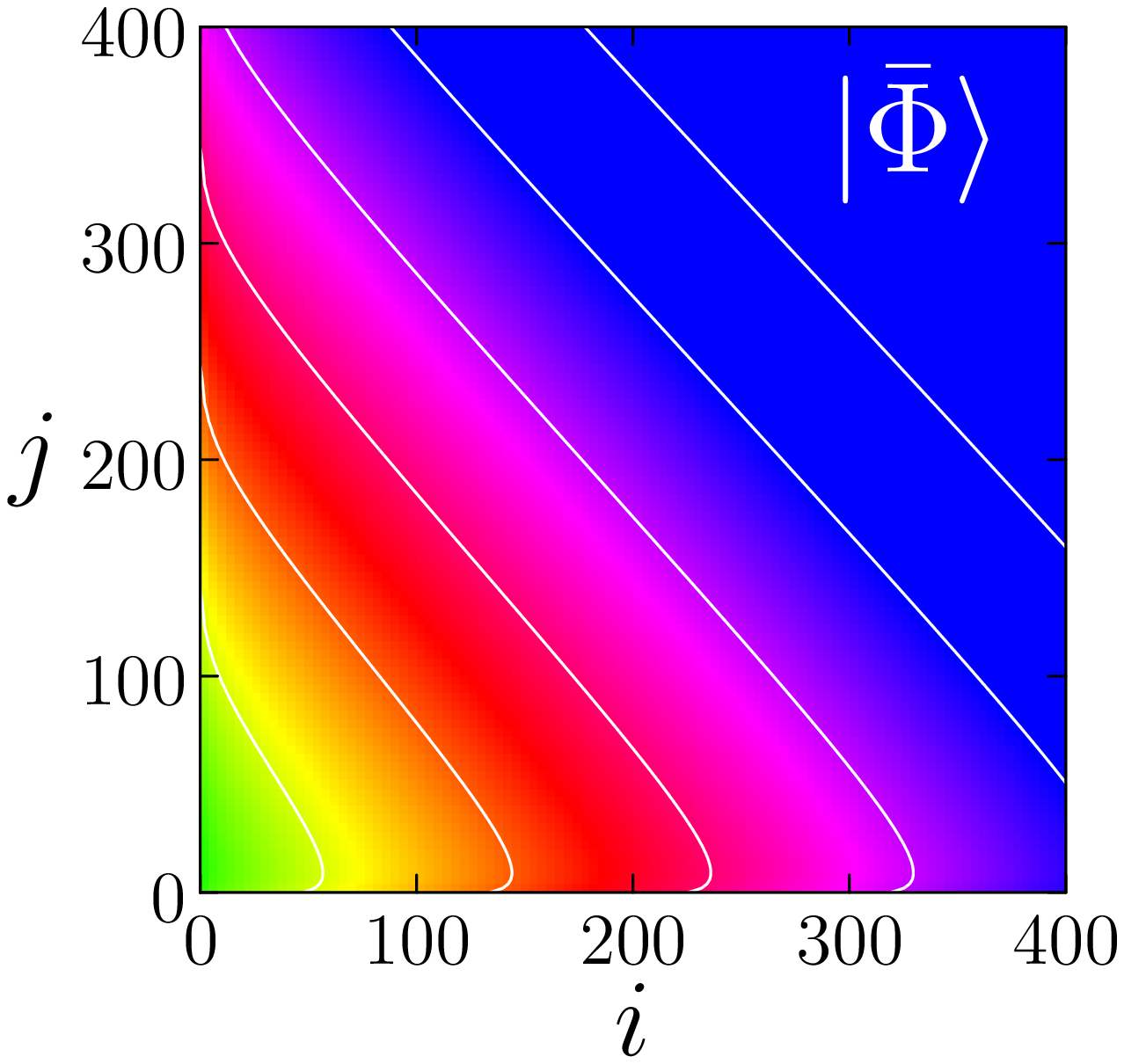}
    \raisebox{2cm}{(b)}\\
    \raisebox{2cm}{(c)}
    \includegraphics[height=3cm]{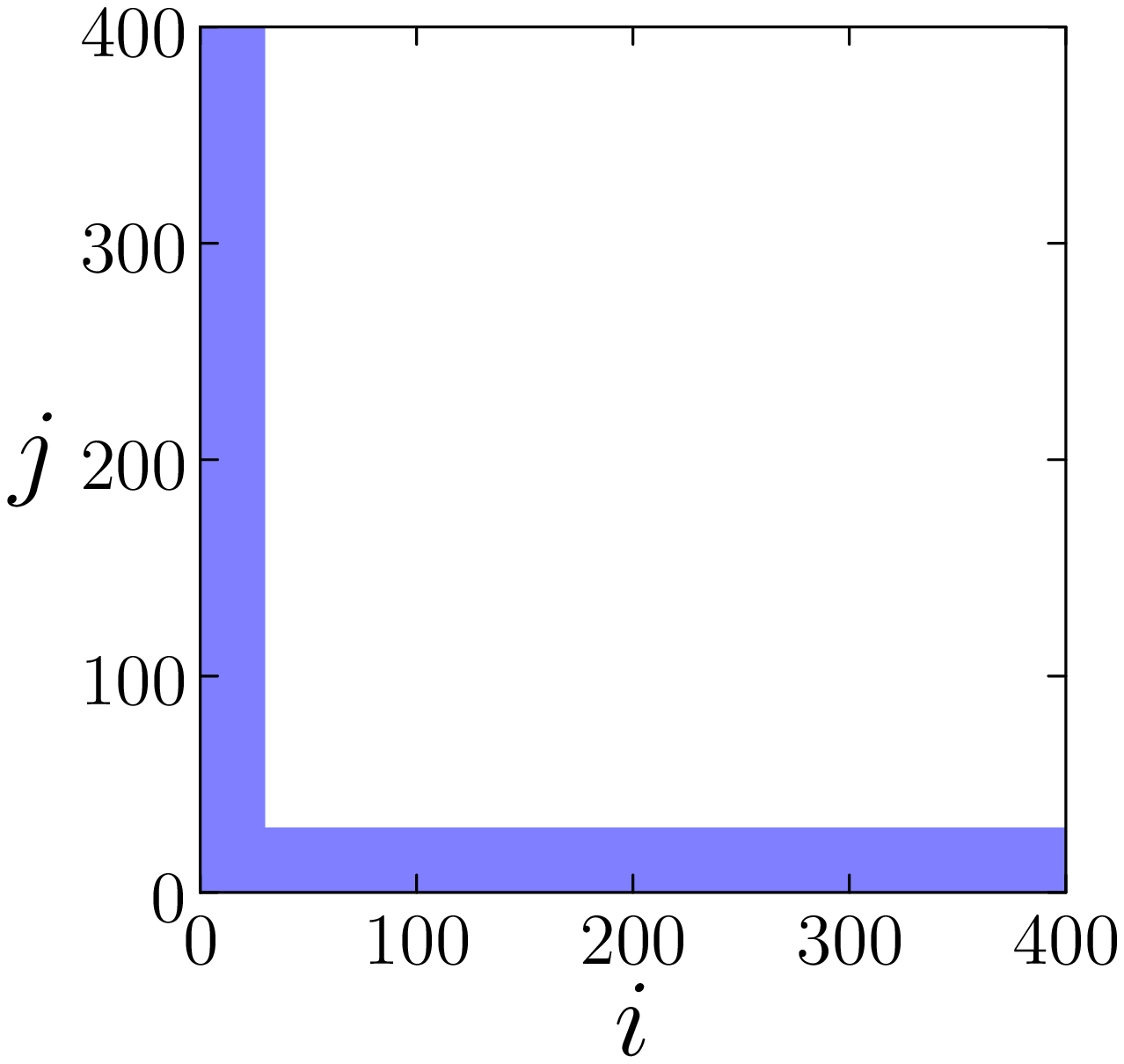}&
    \includegraphics[height=3cm]{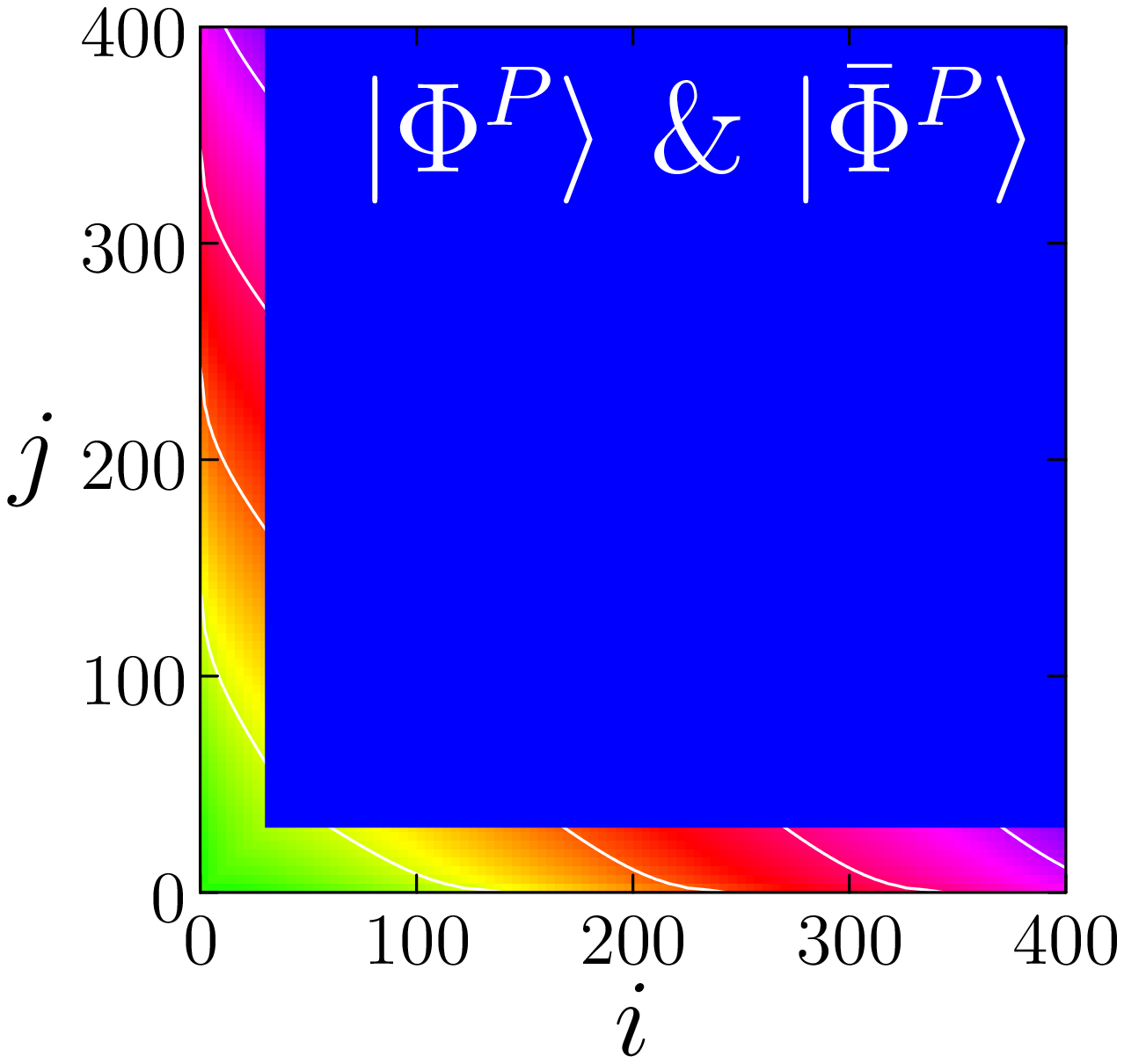}
    \raisebox{2cm}{(d)}\\
    \multicolumn{2}{c}{\includegraphics[height=3cm,angle=-90]{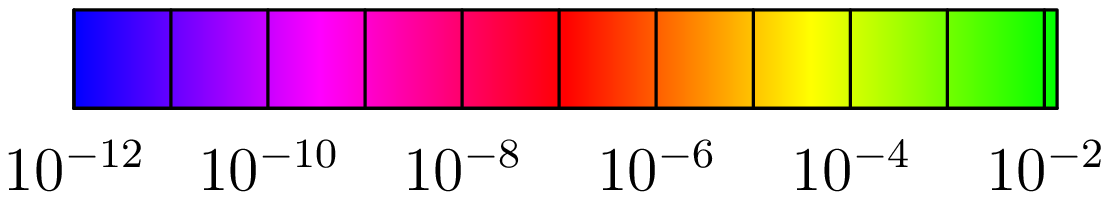}}
  \end{tabular}
  \caption{Photon number distribution of multi-photon states (a) $\lvert\Phi\rangle$, (b) $\lvert\bar{\Phi}\rangle$ before and after preselection (d) $\lvert\Phi^P\rangle$ \& $\lvert\bar{\Phi}^P\rangle$ with the corner filter. The distributions are evaluated for $g=1.1$ and $\delta_{\text{th}}=30$. Action of the filter in the photon number space is shown in (c): only the components of filtered superposition which belong to the blue area are preserved. (Color online)}
    \label{cross}
\end{figure}

We will now examine the preselection procedure $\mathcal{P}_{C}$ for the following preselection condition $\text{C}(\sigma,\Delta)$: $\sigma - \delta_{\text{th}} \le |\Delta|$, where  $\delta_{\text{th}}$ is a threshold. This condition means that those components of the multi-photon superpositions are selected, whose polarization modes population difference is higher than the population sum reduced by $\delta_{\text{th}}$. We call this kind of filtering the corner filter, since the analysis of the modification of a state in terms of its photon number distribution shows that the filter preserves these components of a superposition which belong to the region in the shape of a corner. Photon number distributions for the original multi-photon states and for the ones preselected with the corner filter are depicted in Fig.~\ref{cross}. The filtering is most restrictive if $\delta_{\text{th}}=0$. Here $|\Delta|= \sigma$ and only the N00N-like components are left from the initial polarization singlet. The case of $\delta_{\text{th}} \to \infty$ corresponds to no filtering since $0 \le |\Delta|$ is always fulfilled. 

We will now focus on two cases: $\delta_{\text{th}}=0$ and $\delta_{\text{th}}=2$. Based on the considerations presented at the beginning of the Section IV, we compute the visibility $V^P(\beta)$ , antivisibility $A^P(\beta)$, and the CHSH-Bell parameter $B^P$ for the preselected states in Eq.~(\ref{preselected_multiqubit}). The formulas are lengthy and we display them in Appendix B: Eqs.~(\ref{visibility-preselected-total} - \ref{bell-preselected}).

\begin{figure}
  \begin{center}
    \includegraphics[height=4cm]{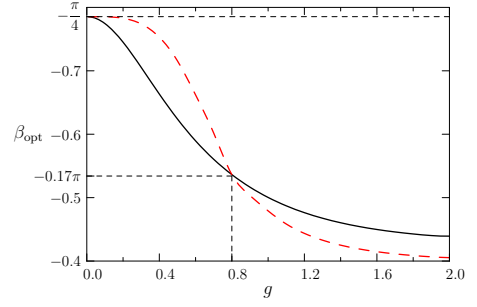}\\
   \end{center}
  \caption{The optimal rotation angle $\beta_{\text{opt}}$ for the multi-photon mode of an unsymmetrical singlet in the CHSH-Bell inequality test, computed for the corner filter with $\delta_{\text{th}}=0$ (black solid line) and $\delta_{\text{th}}=2$ (red dashed line) as a function of amplification gain $g$. (Color online)}     
  \label{fig:beta_gain}
\end{figure}
The optimal rotation angle $\beta_{\text{opt}}$ for the preselected states depends on $\delta_{\text{th}}$ and varies with amplification gain $g$. Fig.~\ref{fig:beta_gain} depicts $\beta_{\text{opt}}$ as a function of $g$ for $\delta_{\text{th}}=0$ and $\delta_{\text{th}}=2$. It can be found numerically by solving $d/d\beta (V(\beta) + A(\beta)) = 0$. The non-orthogonal choice for the measurement settings on the multi-photon mode indicates that the unsymmetrical singlet loses the phase-covariant symmetry after preselection. 

\begin{figure}[ht]
  \begin{flushright}
    \includegraphics[height=4cm]{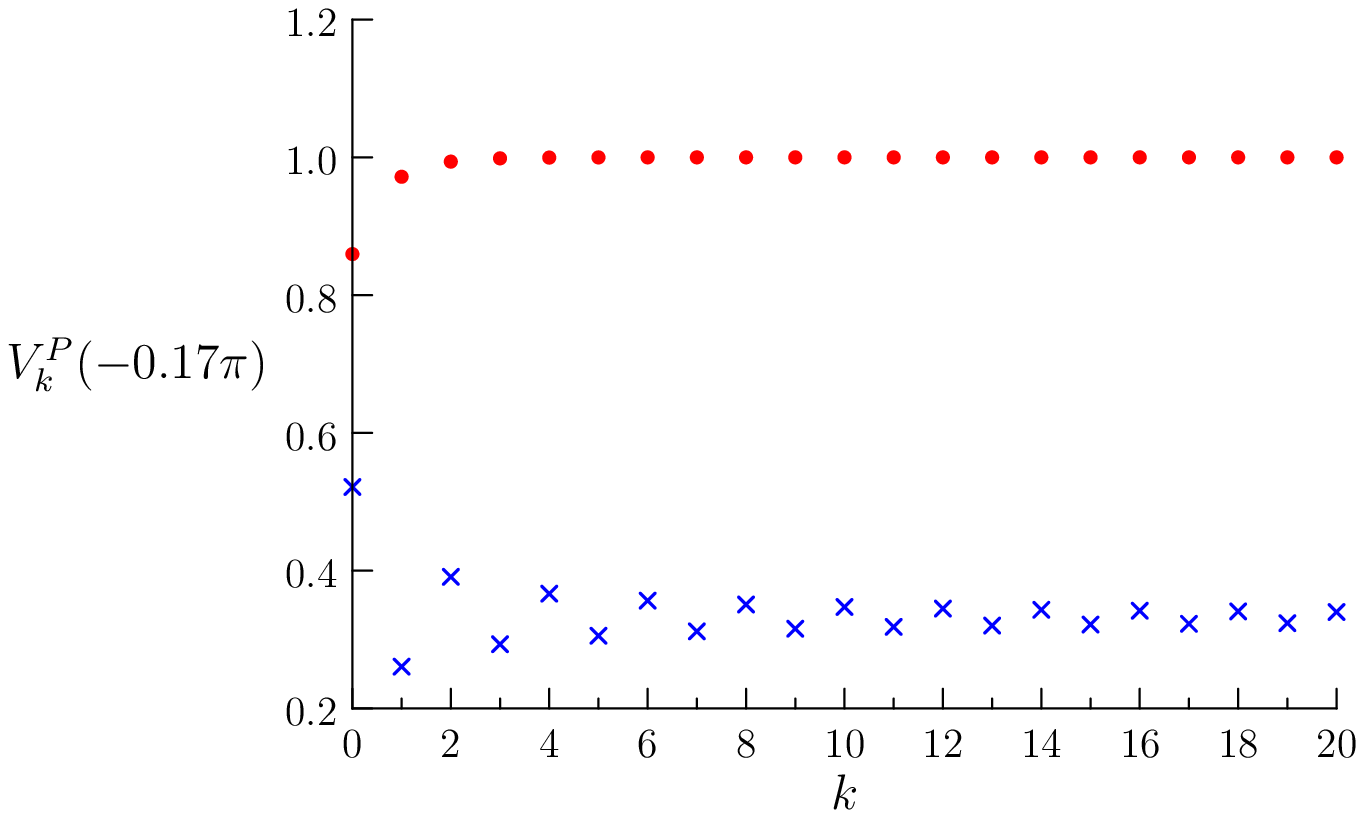}\kern5em\ \\
    \includegraphics[height=4cm]{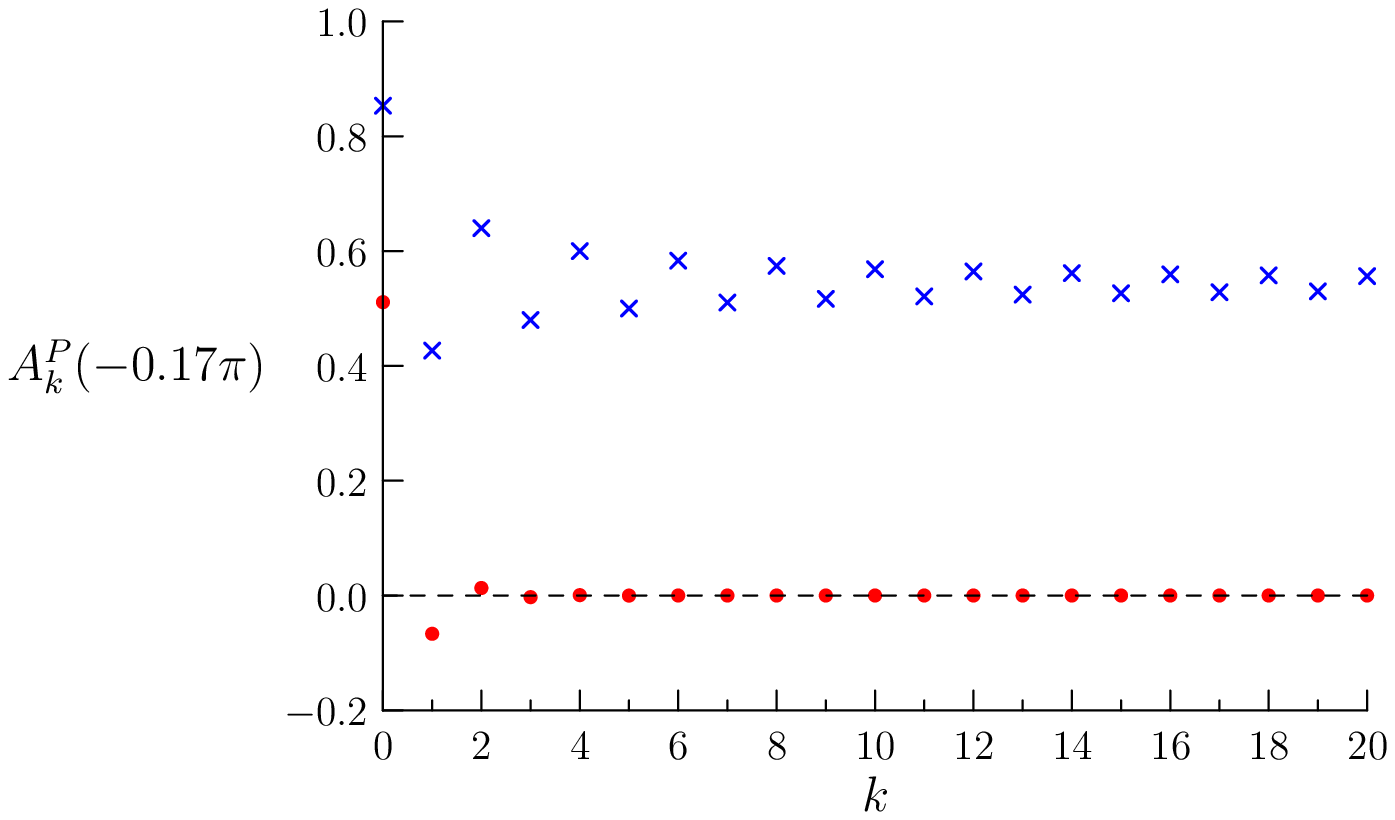}\kern5em\ \\
    \includegraphics[height=4cm]{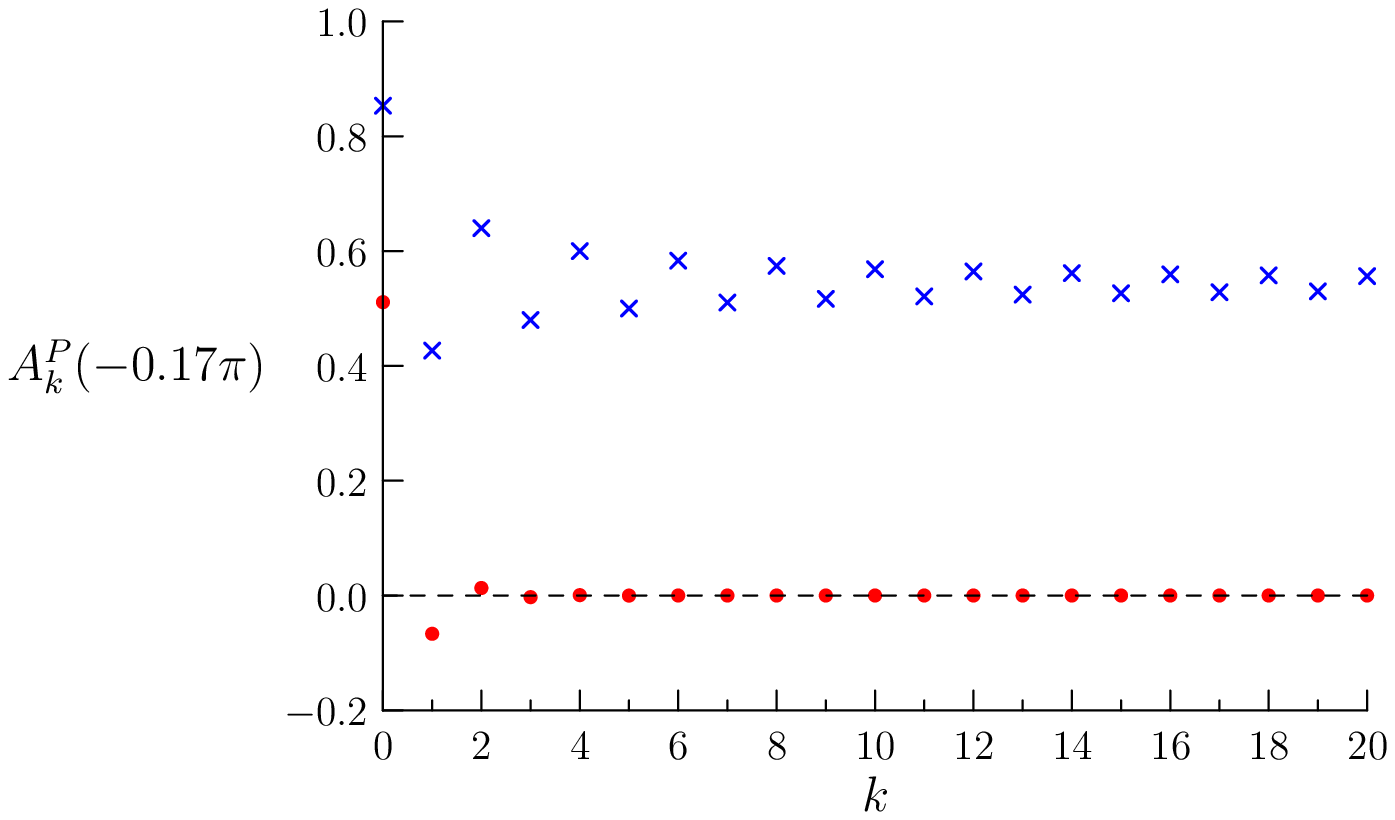}\kern5em\ \\
    \includegraphics[height=4cm]{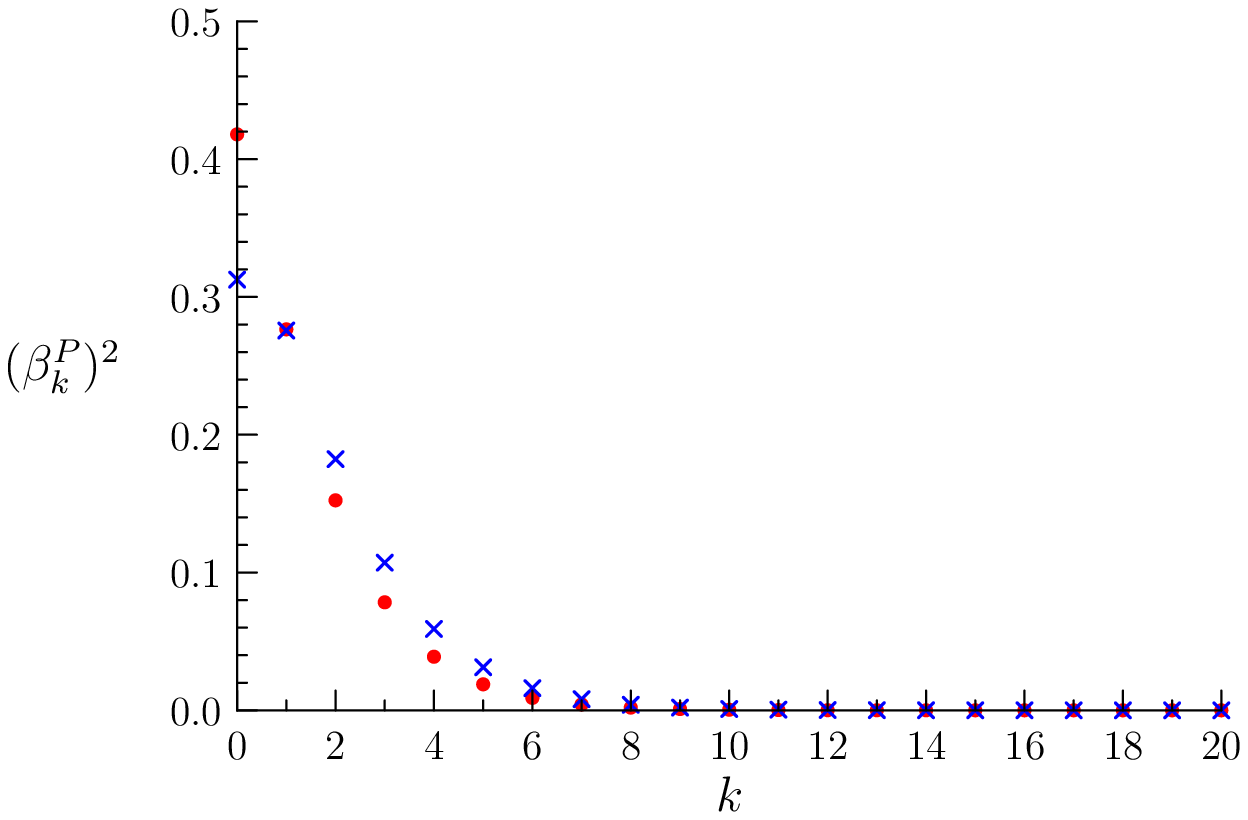}\kern5em\ \\
  \end{flushright}
  \caption{Plot of the visibility $V_k^P$ from Eq.~(\ref{visibility-preselected}) (top figure), antivisibility $A_k^P$ from Eq.~(\ref{antivisibility-preselected}) (upper middle figure), the Bell parameter $B_k^P$ from Eq.~(\ref{bell-preselected}) (bottom middle figure) and $(\beta^P_k)^2$ from Eq.~(\ref{beta}), computed for the multi-photon states $\lvert\Phi^P_k\rangle$ and $\lvert\bar{\Phi}^P_{k}\rangle$ given in Eqs.~(\ref{eq:PhiP})-(\ref{eq:PhiP_bar}) preselected by the corner filter with threshold values $\delta_{\text{th}}=0$ (red dots) and $\delta_{\text{th}}=2$ (blue crosses) as a function of $k$ corresponding to $(2k+1)$-photon-number sector of the multi-photon states. (Color online)}
    \label{fig:afterpreselection}
\end{figure}

Fig.~\ref{fig:afterpreselection} illustrates how the Bell parameter $B^P$ for the states $\lvert\Phi^P\rangle$ and $\lvert\bar{\Phi}^P\rangle$ preselected by the corner filter is constructed. It simultaneously presents the values of: the visibility $V_k^P$ from Eq.~(\ref{visibility-preselected}), antivisibility $A_k^P$ from Eq.~(\ref{antivisibility-preselected}), the Bell parameter $B_k^P$ from Eq.~(\ref{bell-preselected}) and $(\beta^P_k)^2$ from Eq.~(\ref{beta}), computed for the first $k=0,...,20$ sectors (each containing $2k+1$ photons), $g=0.8$ and $\delta_{\text{th}}\in\{0,2\}$. The particular choice of the amplification gain $g$ allows to choose $\beta_{\text{opt}}=-0.17\pi$ for both thresholds (see Fig.~\ref{fig:beta_gain}), making the comparison easier.

In case of $\delta_{\text{th}}=0$ one may notice that $V_k^P=1$ and $A_k^P=0$ for all $k\geq4$, which results in $B_k^P=2$ for sectors of 9 or more photons. Moreover, for $k\in\{0,3\}$ the sum $B^P_k=2(V_k^P+A_k^P)$ is greater than 2 and, together with $(\beta_k^P)^2$ equal to $0.42$ and $0.15$, gives the maximal contribution to the result. In case of $k=1$ the value of $B_k$ is below 2, but due to smaller value of $(\beta^P_{2})^2=0.27$, its negative influence does not completely destroy the total Bell parameter. Finally, $B^P=2.26>B=2.06$ for $g=0.8$ (total mean number of photons equals $4.15$).

Similar behaviour could be observed for other values of $g$ and $\delta_{\text{th}}=0$. Of course $\beta_{\text{opt}}$ changes with $g$, but it is always possible to find $k$ for which $(\beta_k^P)^2$ is relatively large and $B_k^P>2$ at the same time, thus resulting in the total Bell parameter which exceeds the classical limit.

For $\delta_{\text{th}}=2$ in turn, setting rotation angle at Bob's side to $\beta_{\text{opt}}$ gives sawtooth shape of $V^P_k$, $A^P_k$ and thus $B^P_k$.  The values of visibility and antivisibility for the sectors lie between 0 and 1, but nevertheless $B^P_k$ exceeds $2$ for a few $k$, for which $(\beta^P_k)^2$ is the greatest. The obtained Bell parameter equals to $B^P=2.08$, which is less than for $\delta_{\text{th}}=0$ but still a bit more than for not preselected states and the same amplification gain.

The reason behind the shape of $V_k^P$ and $A_k^P$ and thus $B^P_k$ for given filter threshold is the structure of the preselected states obtained with the filter. Setting $\delta_{\text{th}}=0$ allows only the $N00N$-like components of the original polarization singlet to pass through. $\lvert\Phi^P\rangle$ and $\lvert\bar{\Phi}^P\rangle$ become superpositions of $|2i+1,0_{\perp}\rangle$ and $|0, (2i+1)_{\perp}\rangle$ respectively, with varying $i$. Then, the obtained preselected polarization singlet from Eq.~(\ref{micro-macro-singlet1}) takes the new form of a superposition of polarization $N11N$ states with odd $N$
\begin{equation}
|\Psi^-_P\rangle=\sum_{i=0}^{\infty}\tilde{\gamma}_{i0}\big(|1\rangle_A |(2i+1)_{\perp}\rangle_B - |1_{\perp}\rangle_A |2i+1\rangle_B \big),
\label{N11N}
\end{equation}
where $N=2i+1$ and $\tilde{\gamma}_{i0} = \sqrt{\cosh g/2}\gamma_{i0}$.  One can notice that $B^P\geq2$ regardless the gain $g$ for $\delta_{\text{th}}=0$ because for $\beta=0$, which is a suboptimal choice of angle, we have $V^P_k=1$ and $A^P_k=0$.

For $\delta_{\text{th}}=2$, the states $\lvert\Phi^P\rangle$ and $\lvert\bar{\Phi}^P\rangle$ are superpositions not only of $|2i+1,0_{\perp}\rangle$ and $|0,(2i+1)_{\perp}\rangle$ so $N00N$s with odd $N$, but also include terms like $|1,(2j)_{\perp}\rangle$ and $|2j,1_{\perp}\rangle$, i.e. superpositions of $N11N$ states with even $N$. Depending on the parity of $k$, the probability amplitudes are summed up with different signs, resulting in the sawtooth shape of the plot.

\begin{figure}[ht]
  \begin{center}
    \includegraphics[height=2.9cm]{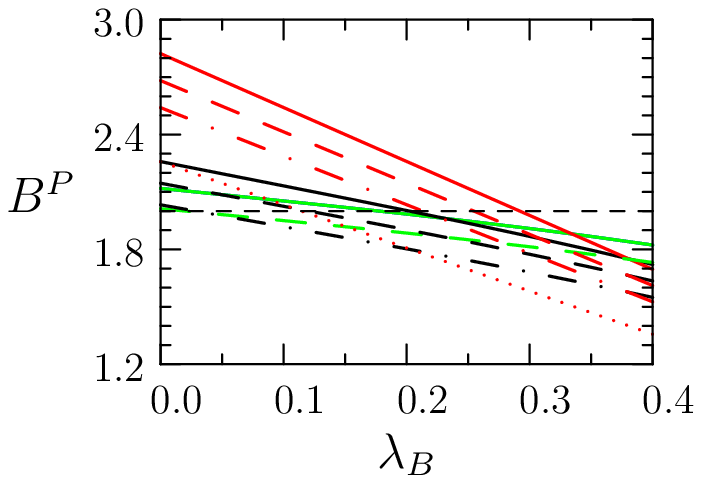}
    \includegraphics[height=2.9cm]{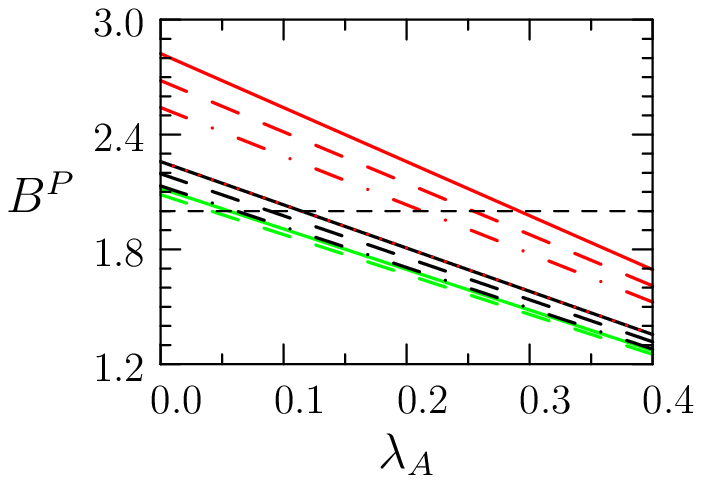}
  \end{center}
  \caption{Bell parameter $B^P$ evaluated for the state $|\Psi^-_P\rangle$ from Eq.~(\ref{N11N}) as a function of losses in the micro-mode $\lambda_A$ and the macro-mode $\lambda_B$ for three values of amplification gain: $g=1.1$ ($8.13$ photons on average, green curves), $g=0.8$ ($4.15$ photons on average, black curves) and $g=0.05$ ($1.01$ photons on average, red curves). The left figure shows $B^P(\lambda_B)$ for fixed values of $\lambda_A=0\%$ (solid line), $\lambda_A=10\%$ (the dashed line),  $\lambda_A=20\%$ (dotted line).  The right figure depicts $B^P(\lambda_A)$ for $\lambda_B=0\%$ (solid line), $\lambda_B=10\%$ (the dashed line),  $\lambda_B=20\%$ (dotted line). (Color online)}
    \label{fig:afterpreselectionLosses}
\end{figure}

At the end of this paragraph we would like to comment on losses for the Bell test with the corner filter as a preselection strategy. As expected, since the filter preselects the state in the N00N-like form, the Bell test is quite fragile to losses in the setup. Fig.~\ref{fig:afterpreselectionLosses} depicts the Bell parameter computed for the state $|\Psi^-_P\rangle$ in Eq.~(\ref{N11N}) for $g=0.05$ ($1.01$ photons on average), $g=0.8$ ($4.15$ photons on average) and $g=0.1$ ($8.13$ photons on average) as a function of losses in the multi- $\lambda_B$ and the single-photon mode $\lambda_A$ for $\delta_{\text{th}}=0$. The violation is much more robust to losses on the amplified-qubit side than on the single photon side. For example, if $\lambda_A=0$ then for gain $g=1.1$ losses on the multi-photon mode up to $20\%$ can be tolerated and $B^P>2$. As expected, the more the state is populated (gain $g$ increases), the less loss-tolerant the Bell test gets.  

\section{Discussion and conclusions}

We have discussed a possibility of performing a postselection-free Bell test with the use of multi-photon quantum states of light and coarse-grained measurements. For this purpose, we examined the CHSH inequality and exemplary unsymmetrical polarization singlet state with the multi-photon states produced in optimal phase covariant quantum cloning. Our work is a proof-of-principle: we show that it is possible to apply a feasible quantum state engineering to multi-photon states and in this way to overcome the problem of imperfect analog detection and violate the classical bound. For the states we discussed, the corner filter is a good choice. It filters out the N00N-like components from the initial superpositions.

We do not claim that the amplified single photons, the CHSH-Bell inequality and the modulus intensity filter are the best strategy for obtaining the loophole-free violation for multi-photon entangled states of light. The CHSH inequality itself imposes the need for the coarse-grained measurements in the case of the unsymmetrical singlets. Perhaps an inequality with much less coarse-graining, i.e. with non-binary measurement outcomes, would be required. Also the analysis of losses in the present model shows that amplification of a two-photon singlet decreases the robustness of the state against losses. Nevertheless, we think that our analysis is an important result because, at least in the near future, it will be difficult to increase arbitralily the resolution of the measuring devices with increasing population of the states, thus to some extent the coarse-graining is unavoidable. Generally, finding a feasible preselection which both enables using coarse-grained detection and creates a state robust against losses, is a difficult task. We conjecture that employing an amplified symmetrical singlet state of light instead of the unsymmetrical one for preselection and Bell test will increase the robustness against losses. 

It is also worth noting that so far there is no proposal allowing for a ``genuine macroscopic'' violation of a Bell inequality. From Figs.~\ref{fig:nopreselection} and \ref{fig:afterpreselection} it is clear that the photon number sectors which contribute to the violation most, come from the small photon numbers. The Bell parameter decreases with increasing photon number $2k+1$. This is a general tendency one observes also for the bright squeezed vacuum state and other Bell inequalities, with observables which are dichotomic or not. Indeed, the preselection helps to increase the values of $B_k$ for all $k$ but the question what is the observable which will reverse the decreasing trend presented in these figures, so that the high photon numbers were contributing to violation most, remains open. 

At the end it is interesting to note that in Ref.~\cite{Tim} it was shown that a Bell inequality violation may be achieved with extremely-coarse-grained measurement in presence of a non-linear interaction. Our results seem to follow this statement: quantum engineering may be viewed as a highly nonlinear operation performed on multi-photon states.

We conclude that taking into account the achievements presented in~\cite{Zeilinger,Kwiat}, it is possible to demonstrate a loophole-free Bell inequality violation for multi-photon singlet states of light within the current technology in the near future.

\acknowledgments

This work is supported by the EU 7FP Marie Curie Career Integration Grant No. 322150 ``QCAT'', NCN grant No. 2012/04/M/ST2/00789, FNP Homing Plus project No. HOMING PLUS/2012-5/12 and MNiSW co-financed international project No. 2586/7.PR/2012/2.  Computations were carried out at the CI TASK in Gda\'nsk and Cyfronet in Krak\'ow.

\section*{Appendix A}

In general case, one may decompose the state $\lvert\Psi\rangle$ into the $k$-photon sectors of the form $\lvert\Phi_k\rangle = \sum_{j=0}^k \xi_{k,j} |k-j, j_{\perp}\rangle$ ($\lvert\bar{\Phi}_k\rangle = \sum_{j=0}^k \bar{\xi}_{k,j} |k-j, j_{\perp}\rangle$ for $\lvert\bar{\Psi}\rangle$), where $\xi_{k,j}$ ($\bar{\xi}_{k,j}$) are certain probability amplitudes. A general form of observable $\mathcal{O}^{(M)}(\beta)$ is given by Eq.~(\ref{eq:general_OM}) and can be expressed as
\begin{align}
\mathcal{O}^{(M)}(\beta) ={}& 
\begin{aligned}[t]
&\sum_{\substack{u,w=0\\C(u,w)}}^{\infty}\sum_{m=0}^u \sum_{n=0}^w (-1)^n \sum_{m'=0}^u \sum_{n'=0}^w (-1)^{n'}\\ &\kern-3em\dfrac{\sqrt{u!\,(m+n)!\,w!\,(u+w-m-n)!}}{m!\,(u-m)!\,n!\,(w-n)!}\\
&\kern-3em\cos^{u-m}\big(\tfrac{\beta}{2}\big) \sin^{m}\big(\tfrac{\beta}{2}\big) \sin^{w-n}\big(\tfrac{\beta}{2}\big) \cos^{n}\big(\tfrac{\beta}{2}\big)\\
&\kern-3em\dfrac{\sqrt{u!\,(m'+n')!\,w!\,(u+w-m'-n')!}}{m'!\,(u-m')!\,n'!\,(w-n')!}\\
&\kern-3em\cos^{u-m'}\big(\tfrac{\beta}{2}\big) \sin^{m'}\big(\tfrac{\beta}{2}\big) \sin^{w-n'}\big(\tfrac{\beta}{2}\big) \cos^{n'}\big(\tfrac{\beta}{2}\big)\\
&\lvert u+w-m-n,(m+n)_\perp\rangle\\
&\langle u+w-m'-n',(m'+n')_\perp\rvert,
\end{aligned}\kern-1em
\end{align}
where $C(u,w)$ represents a condition which ensures the observable to be traceless. The visibility evaluated for this observable equals
\begin{align}
\label{appBeq:Vk}
V_k(\beta)=& \langle\Phi_k\vert\mathcal{O}^{(M)}(\beta)\vert\Phi_k\rangle\\
=&\begin{aligned}[t]
\sum_{\substack{u,w=0\\C(u,w)}}^{\infty}&\Bigg\{\sum_{j=0}^k\xi_{k,j}\sum_{m=0}^u \sum_{n=0}^w (-1)^n \\
&\dfrac{\sqrt{u!\,w!\,(m+n)!\,(u+w-m-n)!}}{m!\,(u-m)!\,n!\,(w-n)!}\\
&\cos^{u-m}\big(\tfrac{\beta}{2}\big) \sin^{m}\big(\tfrac{\beta}{2}\big)\\
&\sin^{w-n}\big(\tfrac{\beta}{2}\big) \cos^{n}\big(\tfrac{\beta}{2}\big)\delta_{j,m+n}\Bigg\}^2.
\end{aligned}\nonumber
\end{align}
Substituting $-\beta$ as a rotation angle in Eq.~(\ref{appBeq:Vk}) gives a formula of a similar form, which differs only with the coefficient $(-1)^j$ in the probability amplitude
\begin{align}
\label{appBeq:Vkm}
V_k(-\beta)
=&\begin{aligned}[t]
\sum_{\substack{u,w=0\\C(u,w)}}^{\infty}&\Bigg\{\sum_{j=0}^k\xi_{k,j}(-1)^j\sum_{m=0}^u \sum_{n=0}^w (-1)^n \\
&\dfrac{\sqrt{u!\,w!\,(m+n)!\,(u+w-m-n)!}}{m!\,(u-m)!\,n!\,(w-n)!}\\
&\cos^{u-m}\big(\tfrac{\beta}{2}\big) \sin^{m}\big(\tfrac{\beta}{2}\big)\\
&\sin^{w-n}\big(\tfrac{\beta}{2}\big) \cos^{n}\big(\tfrac{\beta}{2}\big)\delta_{j,m+n}\Bigg\}^2.
\end{aligned}\kern-1em
\end{align}
The Eqs.~(\ref{appBeq:Vk}) and (\ref{appBeq:Vkm}) are equivalent ($V_k(-\beta)=V_k(\beta)$) when $\xi_{k,j}=\xi_{k,j}(-1)^j$ for all $k$ and $j$.  This is fulfilled when $\xi_{k,j}=0$ for odd $j$.

Similarly, the antivisibility is computed as follows
\begin{align}
\label{appBeq:Ak}
A_k(\beta)=& \langle\Phi_k\vert\mathcal{O}^{(M)}(\beta)\vert\Phi_k\rangle\\
=&\begin{aligned}[t]
&\sum_{\substack{u,w=0\\C(u,w)}}^{\infty}\Bigg\{\sum_{j=0}^k\xi_{k,j}\sum_{m=0}^u \sum_{n=0}^w (-1)^n \\
&\dfrac{\sqrt{u!\,w!\,(m+n)!\,(u+w-m-n)!}}{m!\,(u-m)!\,n!\,(w-n)!}\\
&\cos^{u-m}\big(\tfrac{\beta}{2}\big) \sin^{m}\big(\tfrac{\beta}{2}\big)\\
&\sin^{w-n}\big(\tfrac{\beta}{2}\big) \cos^{n}\big(\tfrac{\beta}{2}\big)\delta_{j,m+n}\Bigg\}\\
&{}\cdot\Bigg\{\sum_{j=0}^k\bar{\xi}_{k,j}\sum_{m=0}^u \sum_{n=0}^w (-1)^n\\
&\dfrac{\sqrt{u!\,w!\,(m+n)!\,(u+w-m-n)!}}{m!\,(u-m)!\,n!\,(w-n)!}\\
&\cos^{u-m}\big(\tfrac{\beta}{2}\big) \sin^{m}\big(\tfrac{\beta}{2}\big)\\
&\sin^{w-n}\big(\tfrac{\beta}{2}\big) \cos^{n}\big(\tfrac{\beta}{2}\big)\delta_{j,m+n}\Bigg\},
\end{aligned}\nonumber
\end{align}
\begin{align}
\label{appBeq:Akm}
A_k(-\beta)
=&\begin{aligned}[t]
&\sum_{\substack{u,w=0\\C(u,w)}}^{\infty}\Bigg\{\sum_{j=0}^k\xi_{k,j}(-1)^j\sum_{m=0}^u \sum_{n=0}^w (-1)^n\\
&\dfrac{\sqrt{u!\,w!\,(m+n)!\,(u+w-m-n)!}}{m!\,(u-m)!\,n!\,(w-n)!}\\
&\cos^{u-m}\big(\tfrac{\beta}{2}\big) \sin^{m}\big(\tfrac{\beta}{2}\big)\\
&\sin^{w-n}\big(\tfrac{\beta}{2}\big) \cos^{n}\big(\tfrac{\beta}{2}\big) \delta_{j,m+n}\Bigg\}\\
&{}\cdot\Bigg\{\sum_{j=0}^k\bar{\xi}_{k,j}(-1)^j\sum_{m=0}^u \sum_{n=0}^w (-1)^n\\
&\dfrac{\sqrt{u!\,w!\,(m+n)!\,(u+w-m-n)!}}{m!\,(u-m)!\,n!\,(w-n)!}\\
&\cos^{u-m}\big(\tfrac{\beta}{2}\big) \sin^{m}\big(\tfrac{\beta}{2}\big)\\
&\sin^{w-n}\big(\tfrac{\beta}{2}\big) \cos^{n}\big(\tfrac{\beta}{2}\big) \delta_{j,m+n}\Bigg\}.
\end{aligned}\kern-1em
\end{align}
The Eqs.~(\ref{appBeq:Ak}) and (\ref{appBeq:Akm}) are equivalent, i.e.\ $A_k(-\beta)=A_k(\beta)$, when for all $k,j$ either $\xi_{k,j}=\xi_{k,j}(-1)^j$ and $\bar{\xi}_{k,j}=\bar{\xi}_{k,j}(-1)^j$ or $\xi_{k,j}=\pm\bar{\xi}_{k,j}(-1)^j$.  Similarly, $A_k(-\beta)=-A_k(\beta)$ when for all $k,j$ $\xi_{k,j}=\pm\xi_{k,j}(-1)^j$ and $\bar{\xi}_{k,j}=\mp\bar{\xi}_{k,j}(-1)^j$.  The last condition is fulfilled e.g. when $\xi_{k,j}=0$ and $\bar{\xi}_{k,j}\not=0$ for odd $j$ but $\xi_{k,j}\not=0$ and $\bar{\xi}_{k,j}=0$ for even $j$.

\section*{Appendix B}

The preselection modifies the sectors of the fixed photon number states in the following way
\begin{align}
  \label{eq:PhiP}
  \lvert\Phi_k^P\rangle=&
  \tfrac{1}{\sqrt{\mathcal{N}_k^P}}\,\mathcal{P}_{\text{C}}\,\lvert\Phi_k\rangle
  \\
  =&\tfrac{1}{\sqrt{\mathcal{N}_k^P}}
  \kern-3em\sum_{\substack{j=0\\\text{C}(\sigma=2k+1,\Delta= 4j+1-2k)}}^k\kern-3em
  \binom{k}{j}\,\sqrt{(2j+1)!\,(2k-2j)!}
  \nonumber\\&
  \qquad\lvert 2j+1,(2k-2j)_{\perp}\rangle,
  \nonumber
\end{align}

\begin{align}
  \label{eq:PhiP_bar}
  \lvert\bar{\Phi}_k^P\rangle=&
  \mathcal{P}_{\text{C}}\,\lvert\bar{\Phi}_k\rangle
  \\
  =&\tfrac{1}{\sqrt{\bar{\mathcal{N}}_k^P}}
  \kern-3em\sum_{\substack{j=0\\\text{C}(\sigma=2k+1,\Delta= 4j-1-2k)}}^k\kern-3em
  \binom{k}{j}\,\sqrt{(2j)!\,(2k+1-2j)!}
  \nonumber\\&
  \qquad\lvert 2j,(2k+1-2j)_{\perp}\rangle,\nonumber
\end{align}
with normalization constants equal
\begin{align}
  \mathcal{N}_k^P=&\kern-3em\sum_{\substack{j=0\\\text{C}(\sigma=2k+1,\Delta= 4j+1-2k)}}^{k}\kern-3em
  \binom{k}{j}^2\,(2j+1)!\,(2k-2j)!,
  \label{norm1}
  \\
  \bar{\mathcal{N}}_k^P=&\kern-3em\sum_{\substack{j=0\\\text{C}(\sigma=2k+1,\Delta= 4j-1-2k)}}^{k}\kern-3em
  \binom{k}{j}^2\,(2j)!\,(2k+1-2j)!.
  \label{norm-bar}
\end{align}
The multi-photon states equal
\begin{align}
  \lvert\Phi^P\rangle=&\sum_{k=0}^{\infty}\beta_k^P\,\lvert\Phi_k^P\rangle,
  \\
  \beta_k^P=& C_g^{-2}\,\left(\dfrac{T_g}{2}\right)^k\,\dfrac{1}{k!}\,
  \sqrt{\dfrac{\mathcal{N}_k^P}{\mathcal{N}^P}},
  \quad
  \sum_{k=0}^{\infty}\left(\beta_k^P\right)^2=1,
  \label{beta}
  \\
  \mathcal{N}^P=& C_g^{-4}\sum_{k=0}^{\infty}\left(\dfrac{T_g}{2}\right)^{2k}\,
  \dfrac{1}{{k!}^2}\,\mathcal{N}_k^P,
  \\
  \lvert\bar{\Phi}^P\rangle=&\sum_{k=0}^{\infty}\bar{\beta}_k^P\lvert\bar{\Phi}_k^P\rangle,
  \\
  \bar{\beta}_k^P=& C_g^{-2}\,\left(\dfrac{T_g}{2}\right)^k\,\dfrac{1}{k!}\,
  \sqrt{\dfrac{\bar{\mathcal{N}}_k^P}{\bar{\mathcal{N}^P}}},
  \quad
  \sum_{k=0}^{\infty}\left(\bar{\beta}_k^P\right)^2=1,
  \\
  \bar{\mathcal{N}}^P=& C_g^{-4}\sum_{k=0}^{\infty}\left(\dfrac{T_g}{2}\right)^{2k}\,
  \dfrac{1}{{k!}^2}\,\bar{\mathcal{N}}_k^P.
\end{align}

\noindent
Let's change the variable $j$ to $j'$ in $\bar{\mathcal{N}}_k^P$ given in Eq.~(\ref{norm-bar}), so
$j'=k-j$ ($j=k-j'$). The sum over $j'$ remains from $0$ to $k$.
\begin{align} \bar{\mathcal{N}}_k^P=&\kern-4em\sum_{\substack{j'=0\\\text{C}(\sigma=2k+1,\Delta=4(k-j')-2k-1)}}^{k}\kern-4em
  \binom{k}{k-j'}^2\,(2(k-j'))!\,(2k+1-2(k-j'))!
  \nonumber\\
  =&
  \kern-4em\sum_{\substack{j'=0\\\text{C}(\sigma=2k+1,\Delta=-(4j'+1-2k))}}^{k}\kern-4em
  \binom{k}{j'}^2\,(2k-2j')!\,(2j'+1)!.
\end{align}
Assuming that $\text{C}(\sigma,\Delta)$ is symmetric with respect to
$\Delta$, we got
$\text{C}(\sigma=2k+1,\Delta=-(4j+1-2k))=\text{C}(\sigma=2k+1,\Delta=4j+1-2k)$,
so $\bar{\mathcal{N}}_k^P=\mathcal{N}_k^P$ and therefore
$\bar{\mathcal{N}}^P=\mathcal{N}^P$ and $\bar{\beta}_k^P=\beta_k^P$.

The observable $\mathcal{O}^{(M)}(\beta)$ is given by Eq.~(\ref{eq:macro_observable}) and can be expressed as
\begin{align}
\mathcal{O}^{(M)}(\beta) ={}& 
\begin{aligned}[t]
&\bigg[\sum_{u-w\ge 0} - \sum_{u-w<0}\bigg]\\
&\kern-4em\sum_{m=0}^u \sum_{n=0}^w \dfrac{\sqrt{u!\,(m+n)!\,w!\,(u+w-m-n)!}}{m!\,(u-m)!\,n!\,(w-n)!}\\
&\kern-3em(-1)^n\cos^{u-m}\big(\tfrac{\beta}{2}\big) \sin^{m}\big(\tfrac{\beta}{2}\big) \sin^{w-n}\big(\tfrac{\beta}{2}\big) \cos^{n}\big(\tfrac{\beta}{2}\big)\\
&\kern-4em\sum_{m'=0}^u \sum_{n'=0}^w \dfrac{\sqrt{u!\,(m'+n')!\,w!\,(u+w-m'-n')!}}{m'!\,(u-m')!\,n'!\,(w-n')!}\\
&\kern-3em(-1)^{n'}\cos^{u-m'}\big(\tfrac{\beta}{2}\big) \sin^{m'}\big(\tfrac{\beta}{2}\big) \sin^{w-n'}\big(\tfrac{\beta}{2}\big) \cos^{n'}\big(\tfrac{\beta}{2}\big)\\
&\lvert u+w-m-n,(m+n)_\perp\rangle\\
&\langle u+w-m'-n',(m'+n')_\perp\rvert.
\end{aligned}\kern-4em
\end{align}

\noindent
Visibility takes the form
\begin{align} 
\label{visibility-preselected-total}
V^P(\beta)=&\sum_{k,k'=0}^{\infty}\beta_k^P\,\beta_{k'}^P\,\langle\Phi_k^P\vert\mathcal{O}^{(M)}(\beta)\vert\Phi_{k'}^P\rangle
 \\ =&\sum_{k}^{\infty}\left(\beta_k^P\right)^2\,\langle\Phi_k^P\vert\mathcal{O}^{(M)}(\beta)\vert\Phi_k^P\rangle
  \nonumber\\
  =&\sum_{k=0}^{\infty}\left(\beta_k^P\right)^2\,V_k^P(\beta)
  \nonumber
\end{align}
where
\begin{align}
\label{visibility-preselected}
V_k^P(\beta)=& \dfrac{1}{\mathcal{N}_k}
\begin{aligned}[t]
&\bigg[\sum_{u-w\ge 0} - \sum_{u-w<0}\bigg]\dfrac{\delta_{u+w,2k+1}}{u!\,w!}\\
&\Bigg\{\kern-3em\sum_{\substack{j=0\\\text{C}(\sigma=2k+1,\Delta= 4j+1-2k)}}^k\kern-3em \binom{k}{j} (2j+1)!\,(2k-2j)!\\
&\quad\sum_{m=0}^u \sum_{n=0}^w \binom{u}{m}\binom{w}{n}(-1)^n\\
&\qquad\cos^{u-m}\big(\tfrac{\beta}{2}\big) \sin^{m}\big(\tfrac{\beta}{2}\big)\\
&\qquad\sin^{w-n}\big(\tfrac{\beta}{2}\big) \cos^{n}\big(\tfrac{\beta}{2}\big)\delta_{2k-2j,m+n}\Bigg\}^2.
\end{aligned}\kern-3em
\end{align}

From the form of Eq.~(\ref{visibility-preselected}) it is possible to derive the property $V^P_k(-\beta)$ = $V^P_k(\beta)$, which implies $V^P(-\beta)$ = $V^P(\beta)$.  This is consistent with the condition found in Appendix~A, since probability amplitudes of $\lvert\Phi_k^P\rangle$ are nonzero only for even number of photons in one of the polarizations. Similarly, the antivisibility equals
\begin{align}
  A^P(\beta)=&\sum_{k,k'=0}^{\infty}\beta_k^P\,\bar{\beta}_{k'}^P\,\langle\Phi_k^P\vert\mathcal{O}^{(M)}(\beta)\vert\bar{\Phi}_{k'}^P\rangle
  \\
  =&\sum_{k}^{\infty}\beta_k^P\,\bar{\beta}_{k}^P\,\langle\Phi_k^P\vert\mathcal{O}^{(M)}(\beta)\vert\bar{\Phi}_{k}^P\rangle
  \nonumber\\
  =&\sum_{k}^{\infty}\left(\beta_k^P\right)^2\,A_k^P(\beta),\nonumber
\end{align}

\allowdisplaybreaks[0]

\noindent
where
\begin{align}
\label{antivisibility-preselected}
A^P_k(\beta)
={}& 
\begin{aligned}[t]
&\dfrac{1}{\mathcal{N}_k}\bigg[\sum_{u-w>0} - \sum_{u-w<0}\bigg]\dfrac{\delta_{u+w,2k+1}}{u!\,w!}\\
&\Bigg\{\kern-3em\sum_{\substack{j=0\\\text{C}(\sigma=2k+1,\Delta= 4j+1-2k)}}^k\kern-3em \binom{k}{j} (2j+1)!\,(2k-2j)!\\
&\quad \sum_{m=0}^u \sum_{n=0}^w \binom{u}{m}\binom{w}{n}(-1)^n\\
&\qquad\cos^{u-m}\big(\tfrac{\beta}{2}\big) \sin^{m}\big(\tfrac{\beta}{2}\big)\\
&\qquad\sin^{w-n}\big(\tfrac{\beta}{2}\big) \cos^{n}\big(\tfrac{\beta}{2}\big)\delta_{2k-2j,m+n}\Bigg\}\\
\cdot&\Bigg\{\kern-3em\sum_{\substack{j=0\\\text{C}(\sigma=2k+1,\Delta= 4j-1-2k)}}^k\kern-3em \binom{k}{j} (2j)!\,(2k+1-2j)!\\
&\quad \sum_{m=0}^u \sum_{n=0}^w \binom{u}{m}\binom{w}{n}(-1)^n\\
&\qquad \cos^{u-m}\big(\tfrac{\beta}{2}\big) \sin^{m}\big(\tfrac{\beta}{2}\big)\\
&\qquad \sin^{w-n}\big(\tfrac{\beta}{2}\big) \cos^{n}\big(\tfrac{\beta}{2}\big)\delta_{2k+1-2j,m+n}\Bigg\},
\end{aligned}\kern-3em
\end{align}
which implies $A^P_k(-\beta)=-A^P_k(\beta)$ and therefore $A^P(-\beta)=-A^P(\beta)$.  Again, it is consistent with the condition derived in Appendix~A, because for the same number of photons in both polarizations, probability amplitudes in $A^P_k(-\beta)$ and $A^P_k(\beta)$ have the same modules and opposite signs. Finally, the Bell parameter (Eq.~(\ref{eq:Bell_any_angles})) can be simplified for angles $\theta=0$, $\alpha=0$, $\alpha'=\tfrac{\pi}{2}$ and $\beta'=-\beta$ to
\begin{align}
  B^P={}&2V^P(\beta)+2A^P(\beta),\nonumber\\
  B={}&2\sum_{k=0}^{\infty}\left(\beta_k^P\right)^2\,V_k^P+2\sum_{k=0}^{\infty}\left(\beta_k^P\right)^2\,A_k^P,
  \nonumber\\
  B^P={}&\sum_{k=0}^{\infty}\left(\beta_k^P\right)^2\,B_k^P,
\end{align}
where
\begin{align}
  B_k^P={}&2\big(V_k^P(\beta)+A_k^P(\beta)\big).
  \label{bell-preselected}
\end{align}

The formulas in Eqs.~(\ref{eq:PhiP})-(\ref{bell-preselected}) hold true also for the Bell test without preselection.  In this case, the condition $C(\sigma,\Delta)$ is always fulfilled and $\lvert\Phi^P\rangle=\lvert\Phi\rangle$, $\lvert\bar{\Phi}^P\rangle=\lvert\bar{\Phi}\rangle$, $\beta^P_k=\beta_k$, $\mathcal{N}^P_k=\mathcal{N}_k$, $\mathcal{N}^P=1$ and $B^P_k=B_k$.


\vfill

\begin{thebibliography}{99}

\bibitem{Arndt2009} M. Arndt, T. Juffmann, an V. Vedral, HFSP Journal {\bf 3}, 386 (2009).

\bibitem{Arndt1999} M. Arndt {\it et al}, Nature {\bf 401}, 680 (1999).

\bibitem{Brezger2002} B. Brezger {\it et al}, \prl {\bf 88}, 100404   (2002).

\bibitem{Mohseni2008} M. Mohseni, P.  Rebentrost, S. Lloyd, and A. Aspuru-Guzik, J. Chem. Phys. {\bf 129}, 174106 (2008).

\bibitem{Zeilinger} M. Giustina {\it et al} Nature {\bf 497}, 227 (2013).

\bibitem{Kwiat} B. G. Christensen {\it et al} \prl {\bf 111}, 130406 (2013).

\bibitem{Bell} J. S. Bell, Physics {\bf 1}, 195 (1965).

\bibitem{CHSH} J. F. Clauser, M. A. Horne, A. Shimony, and R. A. Holt, \prl {\bf 23}, 880 (1969).
  
\bibitem{Acin2007} A. Acin, N. Brunner, N. Gisin, S. Massar, S. Pironio, and V. Scarani, Phys. Rev. Lett. {\bf 98}, 230501 (2007).

\bibitem{Pironio2010} S. Pironio et al., Nature (London) 464, 1021 (2010).

\bibitem{Zukowski} C. Brukner, M. Zukowski, J.-W. Pan, A. Zeilinger, Phys. Rev. Lett. {\bf 92},  127901 (2004).
    
\bibitem{detectors1} A. E. Lita, A. J. Miller, and S. W. Nam, Optics Express {\bf 16}, 3032 (2008).

\bibitem{detectors2} T. Gerrits {\it et al}, Optics Express {\bf 20}, 23798 (2012).

\bibitem{Simon} S. Raeisi, P. Sekatski, and C. Simon, Phys. Rev. Lett. {\bf 107}, 250401 (2011).

\bibitem{DeMartini2008} F. De Martini, F. Sciarrino, and C. Vitelli, \prl {\bf 100}, 253601 (2008).

\bibitem{Macrobell} T. Sh. Iskhakov, M. V. Chekhova, G. O. Rytikov, and G. Leuchs, \prl \textbf{106}, 113602 (2011).

\bibitem{WitnesBSV} M. Stobi\'nska, F. T\"oppel, P. Sekatski, and M. V. Chekhova, \pra {\bf 86}, 022323 (2012).

\bibitem{displacement} P. Sekatski, N. Sangouard, M. Stobi\'nska, F. Bussi\`eres, M. Afzelius and N. Gisin, \pra {\bf 86}, 060301(R) (2012).

\bibitem{displacement-exp} N. Bruno, A. Martin, P. Sekatski, N. Sangouard, R. Thew, and N. Gisin, arXiv:1212.3710.

\bibitem{A} C. Vitelli, N. Spagnolo, L. Toffoli, F. Sciarrino, and F. De Martini, \pra {\bf 81}, 032123 (2010).

\bibitem{B} M. Stobi\'nska, P. Sekatski, A. Buraczewski, N. Gisin, and G. Leuchs, \pra {\bf 84}, 034104 (2011).

\bibitem{Stobinska09} M. Stobi\'nska, P. Horodecki, A. Buraczewski, R. W. Chhajlany, R. Horodecki, G. Leuchs, arXiv:0909.1545.  

\bibitem{Vitelli2010-2} C. Vitelli, N. Spagnolo, F. Sciarrino, and F. De Martini, \pra {\bf 82}, 062319 (2010). 

\bibitem{Sekatski2010} P. Sekatski, B. Sanguinetti, E. Pomarico, N. Gisin, and C. Simon, \pra {\bf 82}, 053814 (2010). 

\bibitem{Pomarico2011} E. Pomarico, B. Sanguinetti, P. Sekatski, H. Zbinden, and N. Gisin, New Journal of Physics {\bf 13}, 063031 (2011). 

\bibitem{Stobinska2011} M. Stobi\'nska, F. T\"oppel, P. Sekatski, A. Buraczewski, M. \. Zukowski, M. V. Chekhova, N. Gisin, and G. Leuchs, \pra {\bf 86}, 063823 (2012).    

\bibitem{NJP2014} K.~Yu.~Spasibko, F.~T\"oppel, T.~Sh.~Iskhakov, M.~Stobińska, M.~V.~Chekhova and G.~Leuchs, New~J.~Phys.\ \textbf{16}, 013025 (2014).
        
\bibitem{DeMartini-PRL} F. De Martini, F.  Sciarrino, and Ch. Vitelli, \prl {\bf 100}, 253601 (2008).

\bibitem{Masha} T. Iskhakov, M. V. Chekhova, and G. Leuchs, \prl {\bf 102}, 183602 (2009).  
  
\bibitem{multiporty} M. Stobi\'nska, W. Laskowski, M. Wie\'sniak, and M. \.Zukowski, \pra {\bf 87}, 053828 (2013).

\bibitem{CPC} A. Buraczewski and M. Stobi\'nska, Comp. Phys. Commun. {\bf 183}, 2245 (2012).
  
\bibitem{SciarrinoWigner} N. Spagnolo, C. Vitelli, T. De Angelis, F. Sciarrino, and F. De Martini, \pra {\bf 80}, 032318 (2009).

\bibitem{Stokes} T. Iskhakov, I. N. Agafonov, M. V. Chekhova, and G. Leuchs, \prl {\bf 109}, 150502 (2012).
    
\bibitem{Tim} H. Jeong, M. Paternostro, and T. C. Ralph, Phys. Rev. Lett. {\bf 102}, 060403 (2009).

\end{thebibliography}
\end{document}